\documentclass[12pt,preprint]{aastex}
\shorttitle{Magnetized Winds Asymptotics}
\shortauthors{Heyvaerts and Norman}

\begin{document}

\title{GLOBAL ASYMPTOTIC SOLUTIONS FOR \\
RELATIVISTIC MHD JETS AND WINDS}

\author{ Jean Heyvaerts } 
\affil{Universit\'e Louis Pasteur, Observatoire de Strasbourg
\altaffilmark{1,4} }
\email{heyvaert@astro.u-strasbg.fr}

\and

\author{Colin Norman }
\affil{Department of Physics and Astronomy, Johns Hopkins University \\
and Space Telescope Science Institute\altaffilmark{2,3}  }
\email{norman@stsci.edu}
\bigskip

 
\altaffiltext{1}{Observatoire, Universit\'e Louis Pasteur,
11 Rue de l'Universit\'e, 67000 Strasbourg, France            }

\altaffiltext{2}{Department of Physics and Astronomy,Homewood campus,
The Johns Hopkins University, 3400 North Charles Street, Baltimore, MD
21218}

\altaffiltext{3}{Space Telescope Science Institute,
3700 San Martin Drive, Baltimore, MD 21218}

\altaffiltext{4}{Visiting Scientist at Space Telescope Science Institute
and Department of Physics and Astronomy, Johns Hopkins University}

\begin{abstract}

We consider relativistic, stationary, axisymmetric, polytropic,
unconfined, perfect MHD winds, assuming their five lagrangian first
integrals to be known. The asymptotic structure consists of
field-regions bordered by boundary layers along the polar axis and at
null surfaces, such as the equatorial plane, which have the structure
of charged column or sheet pinches supported by plasma or magnetic
poloidal pressure.  In each field-region cell, the proper current
(defined here as the ratio of the asymptotic poloidal current to the
asymptotic Lorentz factor) remains constant.  Our solution is given in
the form of matched asymptotic solutions separately valid outside and
inside the boundary layers.  An Hamilton-Jacobi equation, or
equivalently a Grad-Shafranov equation, gives the asymptotic structure
in the field-regions of winds which carry Poynting flux to infinity.
An important consistency relation is found to exist between axial
pressure, axial current and asymptotic Lorentz factor.  We similarly
derive WKB-type analytic solutions for winds which are kinetic-energy
dominated at infinity and whose magnetic surfaces focus to
paraboloids.  The density on the axis in the polar boundary column is
shown to slowly fall off as a negative power of the logarithm of the
distance to the wind source.  The geometry of magnetic surfaces in all
parts of the asymptotic domain, including boundary layers, is
explicitly deduced in terms of the first-integrals.

\end{abstract}

\section{Introduction}
\label{intro}
 
Any stationary axisymmetric non-relativistic, rotating,
magnetized wind will collimate at large distances from the source,
under perfect MHD conditions and polytropic thermodynamics \citep{HN89}. 
\citet{ChiuehLiBegelman} showed that these results hold also for relativistic
winds.  We have recently extended our general analysis 
\citep{HNI} by presenting global analytic
asymptotic solutions for non-relativistic winds, valid from the pole
to the equator, assuming given first integrals. This paper extends
our general analysis to relativistic winds.

Flows which bring no Poynting flux to infinity are called
{\it kinetic winds}. Their magnetic surfaces asymptote 
to paraboloids (\citet{HN89}, \citet{ChiuehLiBegelman}).  If
Poynting flux reaches infinity, the flow is called
a {\it Poynting jet}.  The magnetic surfaces then asymptotically approach
cylinders or conical surfaces in which 
cylindrical ones are nested.  We refer to a magnetic surface as being
{\it asymptotically conical} 
if both $r$ and $z$ approach infinity on this
magnetic surface, such that $(z/r)$ approaches a finite
value. 

We find that the asymptotic structure of relativistic winds consists
of {\it field-regions} where the Lorentz force vanishes in the
direction normal to magnetic surfaces.  These regions are bounded by
null surfaces where the magnetic field vanishes. In the
vicinity of null surfaces, the plasma pressure, 
or possibly the poloidal magnetic pressure, is
significant. The vicinity of the polar axis is also a boundary layer
region.  

Relativistic winds are likely to be present near pulsars, active
galactic nuclei, microquasars and gamma ray burst sources.  Their
structure has been first discussed assuming conical magnetic surfaces
(\citet{Michel69}, \citet{GJ}). 
The cross field force balance has been considered by
\citet{Okamoto75} and \citet{HeinemannOlbert}.  
It has then been recognized
that both non-relativistic and relativistic winds focus asymptotically
(\citet{BlandfordPayne}, \citet{HN89}, \citet{ChiuehLiBegelman})
and that MHD forces may
contribute to the wind acceleration. 
It is possible however that a 
confinement mechanism other than the action of the hoop stress should 
be operating (\citet{Spruitetal97}, \citet{LucekBell97}). 
The general-relativistic dynamics 
of MHD winds for given field geometries has been discussed by
\citet{Camenzind86a, Camenzind86b, Camenzind89} and 
by \citet{Takahashietal90}.
Much work has also been devoted to the
determination of the shape of the magnetic surfaces and to the
wind dynamics.  These studies considered the
special-relativistic (\citet{Ardavan}, \citet{Camenzind87}, 
\citet{LiThesis, Li93b}) 
as well as the general-relativistic transfield equation 
that describes force balance accross field lines (\citet{MobarryLovelace},
\citet{Camenzind87}, \citet{Nittaetal91}, \citet{BeskinParev} 
\citet{Tomimatsu94}, \citet{Koideetal00}).

Analytical models, usually involving some form of self-similarity,
have been presented (\citet{Contopoulos94},
\citet{Contopoulos95}, \citet{ContopoulosLovelace94}, 
\citet{Lovelaceetal91}).
\citet{Lietal92} extended the self-similar analysis
of \citet{BlandfordPayne} to relativistic winds. Their approach, as is
the case here, does not rely on the force-free approximation.  Their
self-similarity requirement restricts the rotation profile of the wind
source to be proportional to $r^{-1}$, imposes the wind source to be
of an infinite extent and the return current flow to be an axial
singularity. This limits the type of asymptotics for this model to
circum-polar current carrying structures. The parabolic shape obtained
by \citet{Lietal92} is made possible by the fact that their wind
source subtends infinite flux (\citet{HNI}, appendix C). By
contrast, their approach deals with the criticality and Alfv\'en
regularity conditions exactly, whereas in our approach it is only
assumed that the set of first integrals is consistent with these
conditions. Because the return current distribution is most important
for determining the degree of collimation, we regard a
non-self-similar approach as preferable for analyzing the asymptotics
of the wind. We also have the possibility of analyzing a point source
of wind which subtends finite total flux.

Solutions for relativistic force-free flows have beeen obtained in
cylindrical symmetry, either for given current profiles (\citet{Appl93a},
\citet{Appl93b}) or for self-consistent plasma flows both
confined (\citet{BeskinMalyshkin00}) or unconfined (\citet{Nitta97}).
\citet{Beskinetal98} construct a solution by expanding in terms
of the inverse of Michel's magnetization parameter.
Cold relativistic winds have been analyzed by \citet{bogovalov99}
for the oblique split-monopole rotator.
The question of the structure and formation of jets 
and their connection to disks have been dealt with numerically in many
papers. Axisymmetric stationary solutions, enforcing
regularity at the light cylinder, have been obtained by \citet{ContopoulosKazanasFendt}. 
Axisymmetric two-dimensional force-free stationary flows
emitted by disks have been calculated by \citet{FendtCamAppl},
\citet{FendtCam96}, \citet{TsinganBogovalov99}
and in three dimensions by \citet{Krasnopolskyetal}, \citet{Koideetal99} and 
\citet{Nishikawaetal97}. The confinement of a star's flow
by the wind from a disk has been studied by \citet{TsinganBogovalov01}.
\citet{VanPutten97}
studied time-dependent simulations of
magnetized relativistic jets. \citet{bogovalov01} finds that
in relativistic winds the collimated region may reduce to a very small
region trapping but little flux. 
Lovelace and collaborators considered the
jet-disk interaction and have used the force free approximation to study the
dynamics and focusing of relativistic jets (\citet{Lovelace76}, 
\citet{Lovelaceetal87}, \citet{Lovelaceetal91},
\citet{Lovelaceetal93}, \citet{Ustyugovaetal00}).
The self-consistent response of the disk emitting the wind has also
been considered in two dimensions (\citet{BellLucek},
\citet{Kudohetal98}, \citet{koide}), and in three dimensions
 \citep{Matsumotoetal96}).  Non-stationary behaviour and an avalanche
 type of accretion flow results.

Special analytical solutions valid in more or less extended regions of
space have been obtained. \citet{LiThesis} and \citet{BegelmanLi94}
give some elements of our general solution below, although without the
complete matching.  In studying the structure of relativistic winds
far from the light cylinder, \citet{Tomimatsu94} adopted a procedure
similar in spirit to our present work.\citet{tomimatsu}, in an elegant
asymptotic analysis, find similar slow logarithmic wind and jet
acceleration. \citet{Nitta95} developed a particular solution in the
limit of winds with a very large mass flux to magnetic flux ratio and
rigid rotation. He matched a solution valid in a cylindrical polar
region, in which a strong current flows, with a conical solution in
the nearby region.  The force balance in the axial region is between
the centrifugal force and the hoop stress, implying that this region
is not much broader than the light cylinder.  Some of our general
asymptotic solutions also have such a mixed structure, although the
force balance near the axis is different. \citet{Okamoto99} insists on
the importance of obtaining solutions consistent with current closure
and has shown that this implies that, in regions where the electric
current returns to the wind source, magnetic lines should bend toward
the equator instead of bending to the axis
\citep{BeskinOkamoto00,Okamoto00}. The solutions presented below,
although their poloidal lines curve to the polar axis in most of the
plasma volume, comply with this requirement.

It is the aim of this paper to provide a general analytical asymptotic
solution for special-relativistic jets, 
assuming the five first integrals of the motion to be known.
It is organized as follows. 
Section (\ref{basics}) recalls the basics of special-relativistic, stationary
axisymmetric rotating MHD winds.  Section
(\ref{secasymptofPoyntingjets}) deals with the field-regions.  The
asymptotic form of the transfield equation in field-regions is
established and reduced to a simple Hamilton-Jacobi equation which we
solve in section (\ref{secsolfield}). 
The solution in the polar boundary layer
is then obtained in section (\ref{secpolarbl}), assuming that it
encompasses little flux.
Our solution applies both to
Poynting jets and to kinetic winds by
means of a WKB approximation.  
We match the field-solution to that which applies 
to the polar boundary layer.  This
specifies how the proper current  
around the polar axis varies with distance to
the wind source. The case of a polar boundary
layer supported by the poloidal
magnetic pressure is also considered
for completeness. In section (\ref{secnullsurf}) we similarly obtain and
match to the field-region a solution valid in the vicinity of a null
magnetic surface, namely the equatorial plane of a magnetic structure
with dipolar type of symmetry. In section (\ref{secshape}) the shape of the
magnetic surfaces is calculated, both in field-regions and in boundary
layers, for both cases of asymptotic regime.  Conclusions regarding
the general properties of relativistic rotating MHD winds are
presented in section (\ref{secconclu}). 

\section{Axisymmetric Stationary Relativistic MHD Flows}
\label{basics}

\subsection{Notation and Definitions}
\label{notations}

We now review relativistic MHD winds and establish our notations. We
use cylindrical coordinates ($r$, $\theta$, $z$).  Unit vectors of the
associated local frame of reference are $\vec{e}_r$,
$\vec{e}_{\theta}$, $\vec{e}_z$. The notation 
$R$ denotes the spherical radius.
The unit normal vector to the poloidal field lines, oriented towards
the polar axis, is $\vec{n}$ and the unit tangent vector to them,
oriented towards increasing $z$, is $\vec{t}$. This vector makes an
angle $\psi$ with its projection on the equatorial plane.
Any vector field can be split into a poloidal
(subscript $P$) and a toroidal (subscript $\theta$) part.
Due to axisymmetry,
the poloidal magnetic field, $\vec{B}_P$, can be expressed in terms of a
magnetic flux function, $a(r,z)$, such that
\begin{equation}
\vec{B}_P = - {1\over r} \ {\partial a \over \partial z} \vec{e}_r
+ {1\over r} \ {\partial a \over \partial r} \vec{e}_z .
\label{Bpversusa}
\end{equation}
A magnetic surface is generated by rotating
a field line about the polar axis. It is 
a surface of constant value of $a(r,z)$.
The magnetic flux through it is  $2 \pi a$.
The flow, described  here in the framework of special relativity, has
a local Lorentz factor, $\gamma$
defined by:
\begin{equation}
\gamma = \left(1 - {{(v_{\theta}^2 + v_P^2)} \over{ c^2}}\right)^{-1/2}
\label{defgamma} \end{equation}
We denote by $\rho$ the proper rest mass density, 
measured in the rest frame of the fluid. The 
momentum per unit proper mass is:
\begin{equation}
\vec{u} = \gamma \vec{v}
\label{specmomentum} \end{equation}

\subsection{First Integrals}
\label{firstintegrals}

A polytropic law is assumed,  by which the proper gas pressure is
related to the proper density by 
\begin{equation}
P = Q(a) \rho^{\Gamma}
\label{polytrop} \end{equation}
where $Q$ is constant following the fluid motion and
$\Gamma$ is the polytropic index. This relation may represent adiabatic or
more complex thermodynamics.
We define the function 
\begin{equation}
\xi = 1 + {\Gamma \over \Gamma -1} {Q \rho^{\Gamma - 1} \over c^2}.
\label{defxi} \end{equation}
which is also equal to $(1+\int dP/\rho c^2)$
calculated at constant polytropic entropy $Q$.
Denoting by $\rho_e$ the electric charge density, by $\vec{j}$ 
the electric current density and by $M_\ast$
the mass of the central object,
the special-relativistic equation of motion can be written
as \citep{GJ,LiThesis}:
\begin{equation}
\gamma \rho (\vec{v} \cdot \vec{\nabla})
(\gamma \xi \vec{v}) =  
-  \vec{\nabla} P + \vec{j} \times \vec{B} + \rho_e \vec{E} 
+ \gamma \rho \vec{\nabla} 
\left(\gamma \xi {{G M_{\ast}}\over{R}}\right)
\label{releqmotion} \end{equation}
The relativistic form of the laws of mass conservation, isorotation,
angular momentum conservation and Bernoulli are obtained as in the
non-relativistic case and involve surface functions $E$, $\alpha$,
$L$, $\Omega$ and $Q$.  The equations
which express these four laws are:
\begin{equation}
\gamma \rho \vec{v}_P = \alpha(a) \vec{B}_P ,
\label{defalpha} \end{equation}
\begin{equation}
\gamma (v_{\theta} - r \Omega(a)) = \alpha(a)  B_{\theta} /\rho ,
\label{defOmega} \end{equation}
\begin{equation}
\gamma \xi r v_{\theta} -{r B_{\theta}\over \mu_0 \alpha(a)} = L(a) ,
\label{defL} \end{equation}
\begin{equation}
\gamma \xi ( c^2 -{G M_{\ast}\over R}) 
- { r \Omega (a)  B_{\theta}\over \mu_0 \alpha(a)}
= E(a) 
\label{defE} \end{equation}
Note that $E$ includes the rest mass energy. The polytropic factor $Q$
is a surface function because, by stationarity, flow surfaces are also
magnetic surfaces.  The rotation rate of the magnetic field,
$\Omega(a)$, which appears in equations (\ref{defOmega}) and
(\ref{defE}) is defined in terms of the electric field by:
\begin{equation}
\vec{E} = - \Omega (a) \vec{\nabla} a
\label{omegaandEfield} \end{equation}
When $\vec{B}_P$ is directed away from the wind source,
$\alpha$ is positive. Since the sense of the magnetic field is immaterial,
it can be assumed that this is so at the positive polar axis. 
For positive $\alpha$, $a$ increases from pole to equator.
We assume $\Omega$ to be always positive.

\subsection{Bernoulli Equation}
\label{BernandTF}

The toroidal variables may be eliminated by using 
Eqs.(\ref{defOmega}) and (\ref{defL}). 
This gives:
\begin{equation}
r B_{\theta} = \mu_0 \alpha \rho \ \ 
{{L -\gamma r^2 \xi \Omega}
\over
{\mu_0 \alpha^2 \xi - \rho}}
\label{Bthetasimple}  \end{equation}
\begin{equation}
\gamma \xi v_{\theta} = {L \over r}
+ {\rho \over r} \ 
{{L -\gamma r^2 \xi \Omega}
\over
{\mu_0 \alpha^2 \xi - \rho}}
\label{vthetasimple}  \end{equation}
We denote by $I$ the quantity
\begin{equation}
I = - {{r B_{\theta} }\over{\mu_0}}
\label{definI} \end{equation}
The minus sign in Eq.(\ref{definI}) has been included to make $I$
positive when $\alpha$ and $\Omega$ are. The physical total poloidal
current is $J_P = - 2 \pi I$. Nevertheless we shall conveniently refer to
$I$ as the poloidal current.  Since $\gamma$ depends on
$v_{\theta}$, the elimination of the toroidal variables in 
Eqs.(\ref{Bthetasimple}) and (\ref{vthetasimple}) is not yet
complete. These expressions can however be substituted in the
Bernoulli equation (\ref{defE}) which can then be solved to obtain an
expression of $\gamma \xi$ in terms of poloidal variables. This
eventually gives the toroidal variables in terms of poloidal ones as:
\begin{equation}
r B_{\theta} = - \mu_0 \alpha
{{ L (c^2 - GM_{\ast}/R) - r^2 \Omega E
}\over{
(c^2 - GM_{\ast}/R) (1 - \mu_0 \alpha^2 \xi/\rho)
-r^2 \Omega^2  }}
\label{Btheta} \end{equation}
\begin{equation}
r v_{\theta} = r^2 \Omega \left(
1 \ \ \ + {{\mu_0 \alpha^2 \xi}\over{\rho}} \ 
{{ r^2 \Omega E -  L (c^2 - GM_{\ast}/R) 
}\over{
 r^2 \Omega E \ (1 - \mu_0 \alpha^2 \xi/\rho) 
- L r^2 \Omega^2 }}\right)
\label{vtheta} \end{equation}
For smooth continuuous solutions, these expressions must be regular where
their denominator vanishes, which implies that when
\begin{equation}
\rho = \mu_0 \alpha^2 \xi \ \ \
 {{c^2 - GM_{\ast}/R 
}\over{
c^2 - GM_{\ast}/R - \Omega^2 r^2 }}
\label{relalfdens} \end{equation}
the position $(r,z)$ must be such that
 \begin{equation}
r^2 = {L \over \Omega} \ \ \
{{(c^2 - GM_{\ast}/\sqrt{r^2 + z^2} )}\over{E}}
\label{relalfpoint}  \end{equation}
Another way to express the special density which appears in
(\ref{relalfdens}) is to insert in it
the expression (\ref{relalfpoint}) for the corresponding radius,
so obtaining a relation between the value assumed by $\rho$ at this 
special point and the first integrals:
\begin{equation}
{{\rho}\over{\xi(\rho)}} = \mu_0 \alpha^2  \ {{E}\over{ E - L \Omega}}
\label{relalfdensofa} \end{equation}
A complete elimination of toroidal variables from Eq.(\ref{defE})
can be achieved as follows. A first expression for $\gamma$
is obtained by substituting Eq.(\ref{Bthetasimple}) in Eq.(\ref{defE}).
An independent relation for $\gamma$ results from 
its definition in Eq.(\ref{defgamma}), using Eqs.(\ref{defalpha}) 
and (\ref{vthetasimple}). Eliminating 
$\gamma$ between these two relations, an equation for
$\rho$, or any other poloidal variable, is obtained.
For given values of the first integrals, this
relation, the relativistic Bernoulli equation, can be used to find this poloidal 
variable as a function of position along the magnetic surface $a$.
Let us denote by $D$ the
variable:
\begin{equation}
D = {{\rho}\over{ \mu_0 \alpha^2 \xi(\rho) }}
\label{defD}
\end{equation}
and by $g$ the function:
\begin{equation}
g = {{G M_*}\over{ c^2 \ \sqrt{r^2 + z^2}  }}
\label{deflittleg} 
\end{equation}
Simple algebra shows that the relativistic Bernoulli 
equation can be then written as:
\begin{eqnarray}
\left( c^2(1 -g) - D (c^2(1-g) - r^2 \Omega^2)\right)^2 
\ \left( \xi^2 + {{B_P^2 }\over{\mu_0^2 \alpha^2 c^2 D^2}} \right) =
\nonumber\\
\left( E - D (E- L \Omega) \right)^2 
+ {{c^2}\over{r^2 \Omega^2}} 
\left( D (E - L \Omega) {{r^2 \Omega^2 }\over{ c^2 }} 
- L \Omega (1 - g) \right)^2
\label{Bernie}
\end{eqnarray}
This equation is satisfied at the
relativistic Alfv\'en point, as can be shown from 
Eqs.(\ref{relalfpoint})-(\ref{relalfdensofa}). Again, 
a smooth solution of Eq.(\ref{Bernie})
requires that the first integrals satisfy regularity conditions 
by passing critical points.

\subsection{Transfield Equation}
\label{subsectfrelat}

The transfield equation is the projection of the equation of motion on
the normal to magnetic surfaces.  It can be obtained  by
using methods similar to those used in 
the classical case \citep{HNI}, giving:
\begin{eqnarray}
{{\alpha \xi} \over{ \rho r}} \left[
{\partial \over \partial z}
{\alpha \over \rho r} {\partial a \over \partial z}
+ {\partial \over \partial r}
{\alpha \over \rho r} {\partial a \over \partial r} \right]
- {1\over \mu_0 \rho r} \left[
{\partial \over \partial z}
{ 1 \over r} {\partial a \over \partial z}
+ {\partial \over \partial r}
{ 1 \over r} {\partial a \over \partial r} \right] \qquad \qquad \qquad &
\nonumber\\
+ {{\Omega^2 }\over{ \mu_0 \rho c^2}}
 \left[ {\partial^2 a \over \partial z^2}+
{1 \over r} {\partial \over \partial r} r {\partial a \over \partial r}
+{\Omega' \over \Omega} \vert \nabla a \vert^2
\right] 
  = 
\gamma E'(a) - {{Q'(a)  \rho^{\Gamma -1}} \over{ \Gamma - 1}}
- {{\Gamma}\over{\Gamma - 1}} {{u_P^2}\over{c^2}} 
{{\vec{\nabla}a \cdot \vec{\nabla}(Q \rho^{\Gamma -1}) 
}\over{
\vert \nabla a \vert^2 }} &
\nonumber\\
+ {\alpha'\over \alpha}{\mu_0 \alpha^2 \rho\over r^2}
\left({L- \gamma \xi r^2 \Omega \over \mu_0 \alpha^2 \xi - \rho}\right)^2
- {{\rho} \over{ r^2 \xi}} (L' - \gamma \xi r^2 \Omega')
\left( {L - \gamma \xi r^2 \Omega \over \mu_0 \alpha^2  \xi - \rho}\right)
- { L L' \over r^2 \xi} \qquad \qquad \qquad &
\label{relTF} \end{eqnarray}
We find that, in comparison of the non-relativistic transfield equation, 
Eq.(\ref{relTF}) has an
additional second order term proportional to $(\Omega^2/c^2)$ on its
left hand side which represents the cross-field component of the
electric force. The gas pressure terms on the right hand side differ
slightly from those obtained by Li (1993), who defined the function
$\xi$ as being $(1 + \int dP/\rho c^2)$.  We prefered the definition
(\ref{defxi}) which coincides with that of Li when the entropy $Q(a)$
is independent of $a$.  Toroidal variables, still implicit in
$\gamma$, can be eliminated entirely by using the expression for
the Lorentz factor in terms of poloidal variables 
(see section \ref{BernandTF}). 
The curvature of poloidal field lines is given by:
\begin{equation}
(\vec{t} \cdot \vec{\nabla}) \vec{t} =
\vec{n} {d \psi \over ds}
\label{fresnet} \end{equation}
Using the relation 
\begin{equation}
(\vec{\nabla} \times \vec{B}_P)\cdot \vec{e}_{\theta} = \vec{n} \cdot \vec{\nabla}
\vert B_P \vert - \vert B_P \vert \ d\psi/ds.
\label{jazandcurvature} \end{equation}
the following equation, equivalent to Eq.(\ref{relTF}), is obtained:
\begin{eqnarray}
\gamma^2 \xi^2
\left(1 - {\rho \over \mu_0 \alpha^2 \xi} \ (1 - {\Omega^2 r^2 \over c^2})  \right)
v_P^2{ d \psi \over ds}
= {{\xi}\over{\rho}} {\vec{\nabla}a \over \vert \nabla a \vert }
\cdot
\vec{\nabla} \left({{\nabla a}^2 \over 2  \mu_0 r^2} + Q \rho^{\Gamma} \right)
\nonumber \\
- \gamma \xi {\vec{\nabla}a \over \vert \nabla a \vert }
\cdot \vec{\nabla} ( \gamma \xi {{ G M_{\ast}}\over{R}} )
- \xi^2  \gamma^2 {v_{\theta}^2 \over  r } {1 \over { \vert \nabla a \vert}}
{\partial a \over \partial r}
+{\xi \over \rho r^2}
{\vec{\nabla}a \over \vert \nabla a \vert }
\cdot \vec{\nabla} \left( {r^2 B_{\theta}^2 \over 2 \mu_0}
-{\Omega^2 r^2 \over c^2} {{\vert \nabla a \vert^2}\over{2 \mu_0}}
\right)
\label{relTFcurvform} \end{eqnarray}
The forces associated with the terms on the right 
hand side are the same as for
classical dynamics, except for the very last one which is part
of the projection of the electric force normal to the magnetic surface.
Another part of this electric force
appears  as a term proportional to $\Omega^2 r^2/ c^2$ 
on the left hand side of Eq.(\ref{relTFcurvform}).

\subsection{Force-free Limit}
\label{subsecforcefreelim}

The force-free limit
applies in the case of very large magnetization. It corresponds
to a limit in which the inertia $\rho$ approaches zero. By Eq.(\ref{defalpha})
this  implies that $\alpha$ approaches zero
such that $\alpha/\rho$ remains finite.
In this limit $\xi$, given by Eq.(\ref{defxi}) 
reduces to unity and
$(\mu_0 \alpha^2 \xi / \rho) \ll 1$. 
Eqs.(\ref{relalfdensofa}),(\ref{defL}),(\ref{defE}) then imply
that the energy is all in Poynting form.
Because the ordering between $\rho$ and the critical density $\mu_0
\alpha^2$ is opposite in the asymptotic limit,
the latter is entirely out of the scope of the force-free approximation. 
The asymptotic limit applies to a state of the
flow reached in regions much further away from the wind source than
the limit down to which the force free approximation is valid.  
Nevertheless, the asymptotic regime is
such that the component of the Lorentz force perpendicular to
magnetic surfaces vanishes, except at boundary layers. There is no
contradiction here \citep{Nitta95}.  This
asymptotic property occurs because the least negligible asymptotic
forces normal to the field are electromagnetic. This property
does not result from any {\it a priori} assumed dominance of
electromagnetic forces over all the other forces present. 
The vanishing of the normal component of the electromagnetic force
simply results from the
wind dynamics.  Between any near-source region in which force free
conditions apply (because of strong magnetization)
and the asymptotic domain (in which the component of the electromagnetic force
perpendicular to the field vanishes as a consequence 
of the dynamics), an intermediate non-force-free region must exist. In this
intermediate non-force-free region, 
currents cross magnetic surfaces and plasma is
accelerated. The conservation of poloidal current on
a magnetic surface (which, under 
force-free conditions, apply near the wind source) 
do not retain validity
continuously out to the asymptotic domain. 

It is instructive to  analyze how the inertia-less limit turns the
transfield equation (\ref{relTF}) into the well known force-free
relativistic wind equation.  In the inertia-less limit,
Eq.(\ref{defL}), multiplied by $\alpha$,
reduces to $ -r B_{\theta} = \mu_0
\alpha L = \mu_0 I(a)$, where, as $\alpha$ approaches
zero, $\alpha L$ approaches the finite limit $I(a)$ which means that
the poloidal current follows magnetic surfaces in the force-free regions and
the associated Lorentz force has no component along the
magnetic field. Multiplying Eq.(\ref{defE}) by $\alpha$ and taking the limit of
vanishing $\alpha$ leads to  $ -r \Omega B_{\theta} = \mu_0 \alpha E
= P$, where it is meant that $\mu_0 \alpha E$ approaches a finite
limit $P$ as $\alpha$ approaches zero. This implies that the energy
flux is all in Poynting form. This relation can be
restated as $ \alpha E = I \Omega$.  The point $r$ defined by
Eq.(\ref{relalfpoint}) reduces in this case to the light cylinder
radius $r = c / \Omega$. In the force-free limit, Eq.(\ref{relTF})
should be expanded in $\rho$ and $\alpha$ to an appropriate
order, to take account of the cancellation of the dominant terms.
The inertial terms
on the left hand side of Eq.(\ref{relTF}) become negligible.
The pressure terms disappear from its right hand side
and $ \alpha L$ approximately equals $I$. One is eventually left
with the well known force-free pulsar wind equation \citep{Beskinbook}:
\begin{equation}
\left( 1 - {{\Omega^2 r^2}\over{c^2}} \right)
\left( {{\partial^2 a}\over{\partial z^2}} + {{\partial^2 a}\over{\partial r^2}} \right)
- \left( 1 + {{\Omega^2 r^2}\over{c^2}} \right)
{1 \over r} \ {{\partial a}\over{\partial r }} - {{r^2 \Omega \Omega'}\over{c^2}} 
\vert \nabla a \vert^2 + \mu_0^2 \ I  I'(a) = 0
\label{pulsareq} \end{equation}


\section {Field Regions}
\label{secasymptofPoyntingjets}

\subsection{Transfield Equation}
\label{subsecasstfrelat}
 
Let us compare the terms on the right of
Eq.(\ref{relTF}) in the large-$r$ limit. 
The gravity term and the centrifugal force term,
which declines as $1/r^3$, become negligible, and the
pressure term becomes very small, so that $\xi$ approaches unity. 
The electric force remains 
of the same order as the hoop stress, though.
The asymptotic form of the relativistic 
transfield equation is then:
\begin{equation}
\gamma^2 v_P^2 
\left(1 + {{\rho r^2 \Omega^2}\over{\mu_0 \alpha^2}} \right){d \psi \over ds}
= {1 \over \rho} {\vec{\nabla}a \over \vert \nabla a \vert }\cdot
\left(
\vec{\nabla} \left({{\nabla a}^2 \over 2  \mu_0 r^2} + Q \rho^{\Gamma} \right) 
+ {1 \over r^2} \vec{\nabla} 
\left({r^2 B_{\theta}^2 \over 2 \mu_0} - {\Omega^2 r^2 \over c^2} 
{{\vert \nabla a \vert^2}\over{2 \mu_0}} \right)  \right)
\label{relassTFcomplete} \end{equation}
Eq.(\ref{defE}) shows that, 
on a given magnetic surface,  $r B_{\theta}$ is bounded at large
distances and  Eqs.(\ref{Bthetasimple}) and (\ref{defalpha})
show that $\rho r^2$ and $r \vert \nabla a \vert$ are also bounded.
When they approach finite limits, the hoop stress term
in Eq.(\ref{relassTFcomplete}) 
decreases as $1/r$, as does the electric force. Both 
the poloidal magnetic pressure and the gas pressure 
decrease more rapidly with $r$. It has been shown \citep{HN89} that 
the inertia force associated with the curvature of the poloidal 
motion on the left of Eq.(\ref{relassTFcomplete}) 
must decrease faster than $1/r$. 
It becomes negligible compared to the hoop stress and electric force.
The  asymptotic form of the transfield equation becomes: 
\begin{equation}
\vec{\nabla}a  \cdot \vec{\nabla}  
\left( r^2 B_{\theta}^2 -  r^2 B_P^2 \ {{\Omega^2 r^2}\over{c^2}}\right) = 0
\label{relassTFoutofbl}\end{equation}
Eq.(\ref{relassTFoutofbl}) generalizes to 
the relativistic case our earlier 
non-relativistic result  
\citep{HN89, HNI,ChiuehLiBegelman}.
Adding to the left hand side of Eq.(\ref{relassTFoutofbl})
that negligible part of the electric force which is
proportional to the curvature of poloidal
field lines the  relation, we find:
\begin{equation}
\vec{\nabla}a \cdot
\left(\vec{j} \times \vec{B} + \rho_e \vec{E} \right) = 0
\label{forcefreeperp}\end{equation}
So, the component of the electromagnetic
force normal to the magnetic surface asymptotically vanishes on 
flared surfaces. This does not imply a strictly force-free situation
since Eq.(\ref{forcefreeperp})
is asymptotically satisfied by the vanishing of each of its terms
separately and holds only perpendicular to
field lines. We refer to regions  where Eq.(\ref{relassTFoutofbl}) holds true 
as field-regions of the asymptotic domain.
The pressure force becomes  significant
near the polar axis and near neutral magnetic surfaces, where
the electromagnetic force vanishes. 
Eq.(\ref{relassTFoutofbl}) can be integrated following orthogonal
trajectories to magnetic surfaces. Let $b$ label one such
orthogonal trajectory. We define $b$ as being the distance to
the origin of the point where this orthogonal trajectory
meets the polar axis.
The integrated form of
Eq.(\ref{relassTFoutofbl}) is:
\begin{equation}
r^2 B_{\theta}^2
-{{\Omega^2 r^2}\over{c^2}}
\vert \nabla a \vert^2  = \mu_0^2 \  K_1(b)
\label{relKassgeneral} \end{equation}
$K_1$ is independent of $a$.
Using Eq.(\ref{omegaandEfield}) and noting that 
$B_{\theta} \gg B_P$ in  the large-$r$ limit, 
Eq.(\ref{relKassgeneral}) can also be written as:
\begin{equation}
r^2 \left(c^2 B^2 - E^2 \right) = \mu_0^2 c^2 \  K_1(b)
\label{fieldinvar2} \end{equation}
The scalar invariant of the electromagnetic field $(c^2B^2 - E^2)$
appears on the left of this equation.
From the asymptotic form of Eqs.(\ref{defalpha}) 
and (\ref{Bthetasimple}), the toroidal field can be expressed 
in terms of poloidal variables as:
\begin{equation}
r B_{\theta} = - {{\gamma_{\infty} \rho r^2 \Omega}\over{\alpha}} =
- {{\alpha}\over {\vert \alpha \vert}} \ 
{{\Omega r \vert \vec{\nabla}a \vert }\over{ v_{\infty} }}
\label{rBthetaass} \end{equation}
When used in Eq.(\ref{relKassgeneral}) this gives:
\begin{equation}
\mu_0^2 K_1(b) = r^2 B_{\theta}^2 \ \left( 1 - \frac{v_{\infty}^2}{c^2} \right)
\label{K2asfunctionofI2} \end{equation}
Then, $K_1(b)$ is positive and can be written as:
\begin{equation}
K_1(b) = K^2(b)
\label{K1K} \end{equation}
$K(b)$ has the dimension of an electric current. 
In the non-relativistic case the quantity which becomes 
independent of $a$ in field-regions is the total poloidal 
electric current $I$. Eq.(\ref{K2asfunctionofI2}) indicates that
the situation  is different in a relativistic 
plasma flow. $K$ and $I$ are related  by
(Chiueh et al. (1991)):
\begin{equation}
I(a,b)  = \gamma_{\infty}(a,b) \ K(b)
\label{IversusK} \end{equation}
This defines $K$ as an algebraic
quantity having  the sign of $I$ 
(see Eqs.(\ref{definI}) and (\ref{rBthetaass})). 
Since the azimuthal velocity asymptotically
approaches zero, 
the terminal Lorentz factor, $\gamma_{\infty}(a)$,
refers to the terminal poloidal velocity $v_{\infty}(a)$. 
The Bernoulli equation (\ref{defE}) becomes in the same limit:
\begin{equation}
\gamma_{\infty} c^2 = E - {{I \Omega}\over{\alpha}}
\label{gammainftyversusEandI} \end{equation}
Eqs.(\ref{IversusK}) and (\ref{gammainftyversusEandI})
provide expressions for the current $I$ 
and the momentum $\gamma_{\infty} v_{\infty}$
in terms of $K(b)$ and the first integrals:
\begin{equation}
I(a,b) = K(b) {{\alpha (a) E(a)}\over{\alpha(a) c^2
+ K(b) \Omega (a) }} \
\label{IversusalphaEonOmega} \end{equation}
\begin{equation}
\gamma_{\infty} v_{\infty} =
c \  {{\sqrt{E^2 -(c^2 +{K \Omega / \alpha})^2 }
}\over{
(c^2 +{K \Omega / \alpha})}}
\label{vinftyversusK} \end{equation}
$K$ is the poloidal current observed in 
a rest frame where the 
poloidal motion vanishes. This can be seen 
by transforming from the laboratory frame to a frame moving with
the fluid at the poloidal velocity $v_{\infty}$ along the direction
of the poloidal field. The azimuthal 
magnetic field is given by Eq.(\ref{rBthetaass}) and the 
electric field by Eq.(\ref{omegaandEfield}).
$K$ is equal to:
\begin{equation}
K =  - {{r B_{\theta} }\over{\mu_0 \gamma_{\infty}}}
= + {{\alpha}\over{\vert \alpha \vert}} \ 
{{ r \Omega \vert \vec{\nabla} a \vert}\over{\mu_0 \gamma_{\infty}
v_{\infty} }}  
\label{KintermsofE}
\end{equation}
In the moving fluid rest frame, the electric field vanishes, while the
azimuthal magnetic field is:
\begin{equation}
B_{\theta, fluid} =  B_{\theta} /\gamma_{\infty} =  - {\mu_0 K \over r}
\label{changeBtheta}
\end{equation}
The arc length element $(r \ d\theta)$ remains invariant in the
transformation, so that:
\begin{equation}
r \ d\theta \ B_{\theta, fluid} = - \mu_0 K d\theta
\label{propercurrent1}
\end{equation}
This shows that $K$ is negatively proportional to
the current enclosed by a circle of radius $r$ 
carried by the fluid motion. We  refer to $K$ as
the proper current.

\subsection{Current-Carrying Boundary Layers and Electric Circuit}
\label{subsecelectcircuit}

Eq.(\ref{relassTFcomplete}) shows that
 the electromagnetic force is proportional 
to $K \vec{\nabla}K$.
Since $K$ vanishes with $I$,
this force yields to pressure at boundary layers
around null surfaces and near the polar axis.
However, since the pressure is weak, the thickness
of the boundary layers must be  small. 
Using Eq.(\ref{rBthetaass}),
Eq.(\ref{relassTFcomplete}) reduces to:
\begin{equation}
{1 \over \rho} (\vec{\nabla} a \cdot \vec{\nabla}) 
 \left(Q \rho^{\Gamma}\right) =  - {{ 1}\over{\rho r^2}}
(\vec{\nabla} a \cdot \vec{\nabla})
\left(  {{r^2 B_{\theta}^{2} }\over{ 2 \mu_0}}
- {{r^2 \Omega^2}\over{ 2 \mu_0 c^2}} \ \vert \nabla a \vert^2
\right) =  - {\Omega \over \alpha} (\vec{\nabla} a \cdot \vec{\nabla})
\left( {{r \Omega \vert \nabla a \vert }\over{ \mu_0 
\gamma_{\infty} v_{\infty} }}\right)
\label{relassTFWKB} \end{equation}
The proper current $K$ then has the following profile
along  an orthogonal trajectory: from zero at the 
 polar axis, it  quickly rises to a non-zero value
at the edge of the circum-polar boundary layer, then stays 
constant and steeply returns to zero through a boundary layer
about the next null surface.
The current system  closes  exactly
in cells  bordered by null surfaces.
Eq.(\ref{bennetequator}) of 
section (\ref{subsecmatchequator}) shows that  $K^2$ resumes its original value
after crossing the boundary layer about a null surface.
$K$ only changes sign.
In the field-region of the next cell, $K$ remains constant, again returning
quickly to zero at the next null surface.

\subsection{Asymptotic Grad Shafranov Equation}
\label{asympgradshafr}

Eq.(\ref{relassTFoutofbl}) can be transformed to give: 
\begin{equation}
{{\partial}\over{\partial z}} \left( {{\alpha}\over{\rho r}}  
{{\partial a}\over{\partial z}} \right) +
{{\partial}\over{\partial r}} \left( {{\alpha}\over{\rho r}}
{{\partial a}\over{\partial r}} \right) +
\vec{n} \cdot \vec{\nabla} 
\left( {{\alpha \vert \vec{\nabla} a \vert}\over{\rho r}} \right) =
{{\alpha \vert \vec{\nabla} a \vert}\over{\rho r}}
{{d \psi}\over{ds}}
\label{identa}
\end{equation}
Since in the asymptotic 
domain $(r \ d\psi/ds)$ becomes negligibly small,
the term on the right of Eq.(\ref{identa}) can be neglected compared to
any one of those on its left. Expanding second order operators
and using Eq.(\ref{defalpha}), this leads, similarly to 
the non-relativistic case,  to the following equation for $a(r,z)$:
\begin{equation}
\Delta a = \vec{\nabla} a \cdot \vec{\nabla} \left( \ln \left( r 
\vert  \nabla a \vert \right) \right) 
\label{curvaturelessfora}
\end{equation}
which can also be restated as:
\begin{equation}
{\rm{div}}\left( {{\vec{n} }\over{r}} \right) = 0
\label{divnoverr}
\end{equation}
Multiplying Eq.(\ref{curvaturelessfora})  by $r^2$ and denoting:
\begin{equation}
{{\vec{\nabla} a \cdot \vec{\nabla} }\over{ \vert \nabla a \vert^2 }}
= {{\partial}\over{\partial a}} 
\label{defpartiala}
\end{equation}
it can also be transformed into:
\begin{equation}
r^2 \ \Delta a = {{\partial}\over{\partial a}}
\left( {{r^2 \vert \nabla a \vert^2}\over{2}} \right) 
\label{pregradshaffass}
\end{equation}
Using Eqs.(\ref{relKassgeneral}), (\ref{rBthetaass}) and (\ref{K1K}),
this can finally be restated as a Grad Shafranov equation:
\begin{equation}
r^2 \ \Delta a = {{\partial}\over{\partial a}}
\left( {{\mu_0^2 K^2(b) \gamma_{\infty}^{2}(a) 
v_{\infty}^{2}(a) }\over{ 2 \Omega^2(a) }} \right) 
\label{gradshaffass}
\end{equation}
The boundary conditions to Eq.(\ref{gradshaffass}) are that $a = 0$ 
along the polar axis and that $a = A$ on the equatorial plane.
When $K(b)$ is constant and non-zero, these conditions imply that
$a$ depends on the latitudinal angle $\psi$ only.
In this case, Eq.(\ref{gradshaffass}) becomes an ordinary differential 
equation for $a(\psi)$ which reduces to the form 
of Eq.(\ref{eqdiffpsiofarel}) below. 

\section{Solution in Field Regions}
\label{secsolfield}

\subsection{Hamilton Jacobi Equation}
\label{subsecHamJacrelat}

Using Eqs.(\ref{rBthetaass}) and (\ref{vinftyversusK}),
Eq.(\ref{relKassgeneral}) can be restated as:
\begin{equation}
r \vert \nabla a \vert 
= {{\mu_0 \vert K(b) \vert \ c 
\sqrt{E^2(a) -(c^2 +{K(b) \Omega(a) / \alpha(a)})^2} }
 \over{
\Omega(a) (c^2 + K(b) \Omega(a) / \alpha(a))  }}
\equiv f(a,b)
\label{preHamJacrel} \end{equation}
Let $\sigma$ be the curvilinear abscissa along
an orthogonal trajectory to magnetic surfaces, conventionnally increasing
from pole to equator. An equivalent form of
Eq.(\ref{preHamJacrel}) is:
\begin{equation}
{{d \sigma}\over{r}} =
{{ (c^2 + K(b) \Omega(a) /\alpha(a)) \ \Omega(a) da 
}\over{ \mu_0 \vert K(b) \vert 
\sqrt{E^2(a) - (c^2 + K(b) \Omega(a) /\alpha(a))^2}  }}
\label{fluxdistcurvilinear}
\end{equation}
When $K(b)$ approaches a finite constant $K_{\infty}$ at large
distances, Eq.(\ref{preHamJacrel}) 
can be further transformed, by defining 
$S(a) = \int_0^a da'/f(a')$, into the following Hamilton-Jacobi equation:
\begin{equation}
\vert \nabla S \vert = {1 \over r}
\label{HamJacobSrel} \end{equation}
the solution of which can be constructed 
by ray tracing, as in the non-relativistic case
\citep{HNI}. The  boundary conditions 
at the equator have been shown to
select a solution in which orthogonal trajectories to magnetic surfaces 
are circles centered at the origin. 
This is only an asymptotic, approximate, result, as is 
Eq.(\ref{HamJacobSrel}) itself.

\subsection{WKBJ Approximation}

When $K(b)$ declines to zero at large distances,
the wind becomes asymptotically kinetic-energy-dominated.
The function $f(a,b)$ of Eq.(\ref{preHamJacrel}) 
asymptotically vanishes in this case. Eq.(\ref{preHamJacrel}) 
then does not give  $r \vert \nabla a \vert$
as a function of  $a$ only. If, however,
the decline of $K(b)$ with distance is very slow,
the WKBJ approximation allows to
neglect this variation.
Orthogonal trajectories then remain
quasi-circular.
In the vicinity of the orthogonal trajectory
of  radius $b$, the flux surfaces are  approximated by a
series of nested conical surfaces
locally represented by
\begin{equation}
z = r  \  \tan (\psi(a, b))
\label{coneeq}
\end{equation}
The angle $\psi(a, b)$ is assumed to slowly  vary with $b$.
 
\subsection{Solution in Field Regions}
\label{subsecsolfieldPoyntrelat}
 
Eq.(\ref{preHamJacrel}) is now considered 
in the WKBJ approximation and
in the geometry described by Eq.(\ref{coneeq}). We find:
\begin{equation}
\vert \nabla a \vert = {{\cos \psi 
}\over{ r \  \vert \partial\psi /\partial a \vert}}
\label{gradaversuspsi} \end{equation}
This gives 
the following differential equation for $\psi$ at given $b$:
\begin{equation}
{{d \psi}\over{cos \psi}}
=
- \ {{ \Omega(a) \  da \  (c^2 + K(b) \Omega(a) /\alpha(a))
  }\over{\mu_0 K(b) \  c \
 \sqrt{E^2(a) - (c^2 +  K(b) \Omega(a)/\alpha(a))^2} }}
\label{eqdiffpsiofarel}\end{equation}
which integrates to:
\begin{equation}
\tan \psi (a,b) = \tan  \psi (a_1,b) 
+ \sinh \left( \int_a^{a_1} {1 \over \mu_0  K  \  c} \
{{ \Omega(a') \  da' \  (c^2 + K \Omega(a')/\alpha(a'))
}\over{
\sqrt{E^2(a') - (c^2 + K \Omega(a') /\alpha(a'))^2 }
}}
\right)
\label{psiofarel}\end{equation}
where $a_1$ is a reference flux in the cell under consideration
and, again, $K$ depends weakly on $b$.
If the cell begins at the equator, $a_1$ can be
taken as the flux variable, $A$,
of the equatorial surface and $\tan(\psi(A,b)) = 0$.
This neglects the small flux in
the equatorial boundary layer, if the latter is a null surface.
These results are similar to those obtained for non-relativistic winds
by \citet{HNI} and for relativistic winds by \citet{Nitta95}.

\subsection{Flux Distribution in Cylindrical Regions of the Field}

When $K(b)$ approaches a finite limit $K_{\infty}$, there 
may exist regions of the free-field where magnetic surfaces become 
cylindrical. Their radius is given by 
Eq.(\ref{fluxdistcurvilinear}), which
in this geometry gives:
\begin{equation}
r_{\infty}(a) = r_{\infty}(a_2)  \ \exp \left(\int_{a_2}^a
{{ \Omega(a') \ da' \  (c^2 + K_{\infty}  \Omega(a') /\alpha(a') )
}\over{
\mu_0 c  K_{\infty}  \
\sqrt{ E^2(a') - (c^2 +  K_{\infty} \Omega(a') /\alpha(a'))^2} }}\right)
\label{rofafarfield} \end{equation}
where $a_2$ is a reference flux in the cylindrical field-region. 
Eqs.(\ref{psiofarel}) and (\ref{rofafarfield}) give 
approximately identical results when $\tan \psi$ becomes very large. 

At this point the solution in the free field is described by 
Eq.(\ref{psiofarel}). To make this 
solution complete, we must solve Eq.(\ref{relassTFWKB}) 
in boundary layers
and determine  how the circumpolar proper current
$K(b)$ depends on $b$.

\section {The Polar Boundary Layer}
\label{secpolarbl}

\subsection{Solution in the  Polar Boundary Layer}
\label{subsecsolpolarbl}

Plasma pressure, or possibly poloidal magnetic pressure,
must be taken into account 
in the vicinity of the polar axis.
Since the poloidal magnetic pressure decreases 
faster with increasing  $r$ 
than the plasma pressure, the latter dominates,
unless the polytropic entropy $Q$ vanishes.
The transfield force balance  near 
the pole is between
the hoop stress, the electric force and 
the pressure gradient. 
Using Eq.(\ref{rBthetaass}), Eq.(\ref{relassTFWKB}) takes the form:
\begin{equation}
(\vec{\nabla} a \cdot \vec{\nabla})
 \left( {{\Gamma}\over{\Gamma -1}} Q \rho^{\Gamma - 1}\right) 
+ {{\Omega}\over{\alpha}} (\vec{\nabla} a \cdot \vec{\nabla})
\left({{\rho r^2 \Omega}\over{\mu_0 \alpha}}\right) = 0
\label{TFinpolarBL}
\end{equation}
Assume that $Q$, $\Omega$ and  $\alpha$
can be taken as constants in the boundary layer. This 
is valid for small $K$ \citep{HNI}. Eq.(\ref{TFinpolarBL}) then 
integrates to:
 \begin{equation}
{{\Gamma }\over{\Gamma - 1}} 
Q_0 \rho^{\Gamma -1}
+ {{\rho r^2 \Omega_0^2 }\over{\mu_0 \alpha_0^2}}
= K_2(b) \label{solpinchcircumpolar} \end{equation}
The integration constant  $ K_2(b)$ can be identified 
by considering the left hand side  of Eq.(\ref{solpinchcircumpolar})
on the polar axis, where the density 
is $\rho_{0}(b)$, so that
 \begin{equation}
K_2(b) = {{\Gamma }\over{\Gamma - 1}}  Q_0 \rho_{0}^{\Gamma -1} (b)
\label{K2ofrho0}
\end{equation}
Alternatively $K_2(b)$ can be identified by considering
the left of Eq.(\ref{solpinchcircumpolar})
at large axial distances, where the pressure becomes negligible. 
Using Eqs.(\ref{Bpversusa}), (\ref{defalpha}), 
(\ref{rBthetaass}) and (\ref{relKassgeneral}) this gives:
\begin{equation}
K_2(b) = {{\Omega_0}\over{\alpha_0}} K(b)
\label{K2versusK}
\end{equation}
where $K(b)$ is the proper current, defined by (Eq.(\ref{K1K}). 
Eq.(\ref{solpinchcircumpolar}) can be solved for $r$ in terms of the 
parameter
\begin{equation}
x = \rho/\rho_0(b)
\label{definexcylindr} \end{equation}
This results in Eq.(\ref{rofxWKB}). Using this expression for $\psi$ in terms
of $x$ in Eq.(\ref{gradaversuspsi})
to express $a$ as a function of $x$, a parametric 
representation of the 
flux distribution in the polar boundary layer is obtained:
\begin{equation}
\cos^2 \psi =   {{\Gamma}\over{\Gamma - 1}} \
{{Q_0 \ \mu_0 \alpha_0^2 \  \rho_0^{\Gamma -2}(b) }\over{
 \Omega_0^2 b^2}}
\left({1 \over x} - {{1}\over{x^{2 - \Gamma}}} \right)
\label{rofxWKB} \end{equation}
\begin{equation}
a = {\gamma}_{0}(b)  v_0(b) {{\Gamma Q_0 \rho_0^{\Gamma -1}(b)}\over{
2 (\Gamma - 1)}}
{{\mu_0 \alpha_0}\over{\Omega_0^2}}
\left(\ln({1 \over x}) 
-  {{2 - \Gamma}\over{\Gamma - 1}} (1 - x^{\Gamma -1}) \right)
\label{aofxrelWKB} \end{equation}
The Lorentz factor and velocity ${\gamma}_{0}(b)$ and $v_0(b)$ 
refer to their values at the polar axis
at a distance $b$ from the source.
Eqs.(\ref{defE}) and (\ref{defxi}) give
${\gamma}_{0}$ as:
\begin{equation}
{\gamma}_0 \ 
\left(c^2 + {\Gamma \over \Gamma -1} 
Q_0 \rho_0^{\Gamma - 1}(b)\right) 
= E_0
\label{gamma0ofE0} \end{equation}
from which  we get 
\begin{equation}
{\gamma}_{0} v_0  = c 
\ {{\sqrt{
E_0^2 -(c^2 + {{\Gamma}\over{\Gamma -1}} Q_0 \rho_0^{\Gamma -1} (b))^2
} }\over{ c^2 + {{\Gamma}\over{\Gamma -1}} Q_0 \rho_0^{\Gamma -1} (b)  }}
\label{gamma0v0} 
\end{equation}

\subsection{Matching the Polar Boundary Layer Solution to the Outer Solution}
\label{subsecmatchpole}

The inner limit ($a \ll A$) of the
outer solution (Eq.(\ref{psiofarel})) must now be 
asymptotically matched
to the outer limit ($x \ll 1$)
of the inner solution, Eqs.(\ref{rofxWKB})--(\ref{aofxrelWKB}).
These equations provide , for $x \ll 1$,
the following expression for  $\psi(a, b)$:
\begin{eqnarray}
\cos^2 \psi(a,b) = {{ \Gamma \  Q_0 \ 
\mu_0 \alpha_0^2 \ \rho_0^{\Gamma -2}
}\over{ (\Gamma - 1) \ \Omega_0^2 b^2}}  
\exp \left( {{ 2 a \ (\Gamma - 1) \ \Omega_0^2 }\over{
\Gamma \ Q_0 \mu_0 \alpha_0 \rho_0^{\Gamma - 1} }}
\ {{ (c^2 + {{\Gamma }\over{\Gamma - 1}} \ Q_0 \rho_0^{\Gamma - 1})
}\over{ c \ 
\sqrt{ E_0^2 - 
(c^2 + {{\Gamma }\over{\Gamma - 1}} \ Q_0 \rho_0^{\Gamma - 1})^2
}  }}  \right)
\label{outerlimofinnersol}
\end{eqnarray}
where $\rho_0$ depends on $b$. 
In the vicinity of the polar axis, $\tan \psi$ is large.
For very small $\cos \psi$, the first term on the right of (Eq. (\ref{psiofarel})) is negligible and this relation can be written as:
\begin{equation}
\cos \psi = {{1}\over{\cosh \chi}} \label{psiandchi} \end{equation}
where the hyperbolic argument $\chi(a, b)$ is:
\begin{equation}
\chi (a, b) = {{1 }\over{\mu_0 c  K(b)  }}
\int_a^{a_1} {{ (c^2 + K(b) \Omega(a') /\alpha(a') ) 
\Omega(a') \ da'}\over{
\sqrt{E^2(a') - (c^2 + K(b) \Omega(a') /\alpha(a'))^2 }   }}
\label{chiinfield} \end{equation}
The apparent weak dependence of $\chi$ on $a_1$ is absorbed by the neglected
first term on the right of Eq.(\ref{psiofarel}). 
The relation (\ref{innerlimofoutersol}) is thus essentially 
independent of $a_1$. For definiteness, $a_1$ can
be taken, in the case of dipolar symmetry, as the equatorial value, $A$, of $a$.
For very small $a$:
\begin{equation}
\chi (a, b) \approx {{1 }\over{\mu_0 c  K  }}
\left( \int_0^{a_1} {{ (c^2 + K \Omega(a') /\alpha(a') ) \ 
\Omega(a') \ da'}\over{
\sqrt{E^2(a') - (c^2 + K \Omega(a') /\alpha(a'))^2 }   }}
- {{ (c^2 + K \Omega_0  /\alpha_0) \ \Omega_0 \ a}\over{
\sqrt{E_0^2  - (c^2 + K \Omega_0  /\alpha_0)^2  }   }}
\right)
\label{chinearpole} 
\end{equation}
From Eqs.(\ref{psiandchi}) and (\ref{chinearpole})
the inner limit of the outer solution 
can be written:
\begin{eqnarray}
{{\cos^2 \psi}\over{4}}  =  \exp\left( - {{2 }\over{\mu_0 c  K }}
\left(
\int_0^{a_1} {{ (c^2 + K\Omega /\alpha ) \ 
\Omega da'}\over{
\sqrt{E^2 - (c^2 + K\Omega /\alpha)^2 }   }}
\ - \ {{ a \ \Omega_0 (c^2 + K \Omega_0  /\alpha_0)}\over{ 
\sqrt{E_0^2  - (c^2 + K \Omega_0  /\alpha_0)^2  }   }} \right)
\right)
\label{innerlimofoutersol}
\end{eqnarray}
For brevity the dependence of $K$ on $b$
and of the first integrals on the integration 
variable $a'$ has been omitted.

\subsection{Bennet Pinch Relation}
\label{subsecbennet}

For the solutions (\ref{outerlimofinnersol}) and
(\ref{innerlimofoutersol}) to smoothly match, it is necessary that the
arguments of their exponential functions of $a$ coincide. This
implies:
\begin{equation}
{{\Gamma }\over{\Gamma - 1}} Q_0 \rho_0^{\Gamma -1}(b) 
= K(b) \ {{\Omega_0  }\over{\alpha_0}}
\label{BennetrelatK} \end{equation}
This is again Eqs.(\ref{K2ofrho0})--(\ref{K2versusK}).
Eq.(\ref{BennetrelatK}) is 
a Bennet pinch relation 
between the proper current 
and the plasma pressure on the axis. The proper current  
$K$ is related by Eq.(\ref{IversusK}) to the asymptotic 
poloidal current and Lorentz factor. 

\subsection{Polar Boundary Layer Proper Current, Density and Radius}
\label{subsecbootstrap}

For a smooth asymptotic matching of 
Eqs.(\ref{outerlimofinnersol}) and (\ref{innerlimofoutersol})
the factors in front of their exponential functions 
of $a$ must also coincide. Let us note 
\begin{equation}
s^2 = {{ \Gamma}\over{ (\Gamma - 1)}} \ Q_0 \rho_0^{\Gamma -1}
\label{sound}
\end{equation}
The velocity $s$ is of order of the sound speed at the axis. 
It depends on $\rho_0(b)$. Taking Eq.(\ref{BennetrelatK}) 
into account, this matching  condition can be written as:
\begin{eqnarray}
{{ \Gamma \ Q_0 \mu_0 \alpha_0^2 \rho_0^{\Gamma -2}
}\over{ (\Gamma - 1) \ \Omega_0^2 b^2}}
= 
4 \ \exp\left( - {{2 \Omega_0}\over{ c \mu_0 \alpha_0  s^2 }}
\int_0^{a_1} {{ (c^2 +  s^2 \
{{\Omega(a') \alpha_0 }\over{\alpha(a')\Omega_0}}) \ 
\Omega(a') \ da'}\over{
\sqrt{E^2(a') - 
(c^2 +  s^2 \ {{\Omega(a') \alpha_0 }\over{\alpha(a')\Omega_0}})^2 }   }}
\right)
\label{equaforrho0}
\end{eqnarray}
Eq.(\ref{equaforrho0}) determines $ \rho_0(b)$
(on which $s^2$ depends by Eq.(\ref{sound})). 
Let us define  a length 
$\ell$, a dimensionless measure of the axial density, $n_0$, 
and a reference magnetic flux $a_0$ by:
\begin{equation}
\ell^2 =  {{\Gamma}\over{\Gamma - 1}}
{{Q_0 (\mu_0 \alpha_0^2)^{\Gamma - 1}}\over{\Omega_0^2}}
\label{scalepinchalf} \end{equation}
\begin{equation}
n_0(b) = {{\rho_0(b) }\over{\mu_0 \alpha_0^2}}
\label{normrhotoalfdens} \end{equation}
\begin{equation}
a_0 = {1 \over 2} \ c \ \mu_0 \alpha_0 \ell^2
\label{referenceflux}
\end{equation}
Let us also introduce the notation
\begin{equation}
\lambda = \int_0^{a_1} 
{{ (c^2 +  s^2 \ 
{{\Omega(a') \alpha_0 }\over{\alpha(a')\Omega_0}})
}\over{
c \ \sqrt{E^2(a') -
(c^2 + s^2 \ {{\Omega(a') \alpha_0 }\over{\alpha(a')\Omega_0}})^2 }}}
\ {{ \Omega(a') \ da'}\over{\Omega_0 \ a_0}}
\label{deflambda}
\end{equation}
The integral $\lambda$ depends on $n_0(b)$, since, by Eq.(\ref{sound}), 
$s^2$ does. The logarithm of Eq.(\ref{equaforrho0})
provides the following equation for $n_0(b)$:
\begin{equation}
{{\lambda(n_0(b))}\over{n_0^{\Gamma -1}(b) }}
= (2 - \Gamma) \ln \left( n_0(b) \right) 
+ \ln \left( {{4 b^2}\over{\ell^2}} \right) 
\label{equaforn0}
\end{equation}
Since $n_0(b)$ is small and ($b/\ell$) large, a
solution can be obtained by iteration, giving, 
in the simplest degree of approximation:
\begin{equation}
n_0^{\Gamma -1}(b) =
{{\lambda((n_0(b)) }\over{ 2 \ln \left( {{2b}\over{\ell}}\right) }}
\label{nzeroofRsimple} \end{equation}
The solutions of Eqs.(\ref{equaforn0}) or (\ref{nzeroofRsimple})
are controlled by the growth 
of the logarithmic term on their right.
They can be satisfied
for large $b$'s in two different ways, according to whether the
proper current $K(b)$ approaches a finite value or decreases to zero.
If $K(b)$ approaches a finite value, Eq.(\ref{BennetrelatK}) 
indicates that the 
axial density becomes  independent of distance. The logarithmic
term in the denominator of equation (\ref{nzeroofRsimple})
must then be compensated by a divergence 
of the numerator term, $\lambda(n_0(b))$.
As $b$ increases,
$K(b)$, related to $n_0(b)$ by Eq.(\ref{BennetrelatK},
should approach a limit that causes the integral
$\lambda(n_0)$ to diverge. 
This implies that $K$ should approach
the flat absolute minimum, $K_{sup}$, of the
function $(\alpha (E - c^2)/\Omega)$. 
If this function does not have such a minimum, 
a solution with a finite limit for $K(b)$ cannot be found and
$K(b)$ should actually approach zero.
Because the square root at the right of
Eq.(\ref{deflambda}) must remain definite for any $a$,
$K_{sup}$ can only be approached from below.
When $K = K_{sup}$, all the energy flux is in Poynting form. 

If $K(b)$ declines to zero at large distances,
$\lambda$ approaches a limit $\lambda_0$
independent of $b$.
Eq.(\ref{nzeroofRsimple}) then shows that $\rho_0(b)$
scales as
$\left({\ln ({{b}/{\ell}}})\right)^{-{{1}\over{\Gamma-1}}}$
and $K(b)$, given by Eq.(\ref{BennetrelatK}),
slowly decreases as 
$\left({\ln ({{b}/{\ell}}})\right)^{-1}$:
\begin{equation}
K(b) =
{{\Omega_0 \ c}\over{\mu_0 \sqrt{E_0^2 - c^4}}}
\left(\int_0^A
da'\  {{\Omega (a') \sqrt{ E_0^2 -c^4} }\over{\Omega_0  \sqrt{E^2(a')  - c^4}
 }}
\right) \ 
{{1}\over{\ln \left({b \over \ell}\right)}}
\label{KofRexplicit} \end{equation}
This slow decrease of $K(b)$ justifies
{\it a-posteriori} the adopted WKBJ procedure.
The hyperbolic argument $\chi$ of Eq.(\ref{chiinfield})
becomes, in this case (taking $a_1$ equal to 
the equatorial value of the flux, $A$):
\begin{equation}
\chi(a,b) =
\int_a^A {{c \ \Omega(a') \  da'}\over{\mu_0 K(b) \sqrt{E^2(a') -c^4}}}
=
k(a) \  \ln \left(b \over \ell \right)
\label{chiofaandR} \end{equation}
where
\begin{equation}
k(a) = {{
\int_a^A da' \
{{\Omega (a') \sqrt{ E_0^2 -c^4}  }\over{\Omega_0  \sqrt{E^2(a')  - c^4}  }}
}\over{
\int_0^A da' \
{{\Omega (a')  \sqrt{ E_0^2 -c^4} }\over{\Omega_0   \sqrt{E^2(a')  - c^4}  }}
}}
\label{kofa} \end{equation}
The solution (\ref{rofxWKB}) shows that the core radius of the jet
at a distance $b$ from the source is:
\begin{equation}
r_0^2(b) = {{\Gamma }\over{\Gamma - 1}} \
{{Q_0 \mu_0 \alpha_0^2 \rho_0^{\Gamma -2}}(b)\over{\Omega_0^2
 }}
\label{pinchradius} \end{equation}
At this point the solution near the pole and in the
field region extending in the polar-most cell is  completely
determined.

It is important to note the very slow decline 
of the proper current and associated Poynting flux that 
Eq.(\ref{KofRexplicit}) implies. The ratio $\sigma$
of the Poynting flux to the other forms of energy flux
is given by Eq.(\ref{defE}):
\begin{equation}
\sigma = - {{r \Omega (a) B_{\theta} 
}\over{ 
\mu_0 \alpha \ (E(a) +  r \Omega (a) B_{\theta} /\mu_0 \alpha ) }}
\label{sigmageneral} \end{equation}
From Eqs.(\ref{definI}) and (\ref{IversusalphaEonOmega}) it results that
\begin{equation}
\sigma = {{K(b) \Omega (a) }\over{ \alpha (a) c^2}} 
\label{sigmaofK} \end{equation}
Eq.(\ref{KofRexplicit}) then explicitly indicates 
how $\sigma$ decreases in a kinetic wind with distance 
$b$ to the wind source, i.e.:
\begin{equation}
\sigma = {{ A \ \Omega_0 \ \Omega (a) }\over{ 
\mu_0 \ c \ \alpha (a) \  \sqrt{E_0^2 - c^4} }} 
\left( \int_0^A {{da'}\over{A}} 
{{ \Omega (a') \sqrt{E_0^2 - c^4} }\over{
\Omega_0 \sqrt{ E^2(a') - c^4} }} \right)
\left( {{1 }\over{ \ln \left( {b / \ell } \right) }} \right)
\label{sigmaexplicitkinetic}
\end{equation}
the convergence of the flow to a completely kinetic energy dominated
state is obviously very slow. \citet{Chiuehetal98}
stress the fact that such a slow decline is a difficulty
for understanding the Crab pulsar wind, which has a 
very large terminal Lorentz factor  and a $\sigma$-parameter as low as
10$^{-2}$ - 10$^{-3}$. We come back in paper III of this series
\citep{HNIII}  on the implications of such a slow decline and
show that, even though the mathematical asymptotics implies a very low
$\sigma$, actual winds may not have converged to this state during
their finite life time, so that $\sigma$ could remain finite when they
reach, for example, a terminating shock.  We do not propose in this
paper any specific solution to the question raised by
\citet{Chiuehetal98} concerning the properties of the Crab pulsar
wind. Such an explanation may have to be found, as
\citet{Chiuehetal98} suggest, in non-ideal effects close to the light
cylinder not considered here.

\subsection{Force-Free Polar Boundary Layers}
\label{subsecffpolarbl}

In the exceptional case when the poloidal magnetic pressure dominates 
over plasma pressure in the asymptotic circumpolar region,
the transfield mechanical balance equation
takes the form of Eq.(\ref{relassTFcomplete}), neglecting the poloidal curvature
term and pressure. In cylindrical geometry, this gives 
\citep{Appl93a,Appl93b}:
\begin{equation}
{{d B_P^2}\over{dr}}  + { 1 \over r^2} \
{{d }\over{dr}}  \ \left( r^2 B_{\theta}^2 -
{{\Omega^2 r^4}\over{c^2}} B_P^2 \right) = 0
\label{pinchrelatff} \end{equation}
Under negligible gas pressure, $\xi =1$.
Eqs.(\ref{defalpha}), (\ref{defE}) and (\ref{Bthetasimple})
provide $B_P$ in terms $B_{\theta}$:
\begin{equation}
B_P^2 = {{c^2 B_{\theta}^2}\over{ r^2 \Omega^2}} \
\left(
1 \ - {{c^4}\over{ (E + r \Omega B_{\theta} /\mu_0 \alpha)^2 }}
\right)
\label{BpolofBtor}  \end{equation}
Eq.(\ref{pinchrelatff}) can then be restated as an equation
for $I$ defined by Eq.(\ref{definI}):
\begin{equation}
{{d }\over{dr}}
\left(
{{I^2}\over{r^4 \Omega^2}}
\left( 1  - {{c^4}\over{ (E - I \Omega  / \alpha)^2}} \right)    \right)
+ {1 \over r^2} \ {{d }\over{dr}}
\left(
{{ I^2 c^2}\over{(E - I \Omega  / \alpha)^2}} \right) = 0
\label{bigeqforIff}  \end{equation}
Assuming the first integrals to be 
constant in the polar boundary layer, which implies 
that $I \ll \alpha E/ \Omega$  \citep{HNI}, 
Eq.(\ref{bigeqforIff}) simplifies to:
\begin{equation}
{{E_0^2 - c^4}\over{c^2 \Omega_0^2}} {{d }\over{dr}}  \left({{I^2}\over{r^4}}
\right)
+ {1 \over r^2} \ {{d I^2}\over{dr}} = 0
\label{eqforIffsimple} \end{equation}
the solution of which is:
\begin{equation}
I = I_{\infty}(b) \ {{c^2 \Omega_0^2 r^2}\over{
c^2 \Omega_0^2 r^2 + (E_0^2 - c^4) }}
\label{solffpinchrelat} \end{equation}
The characteristic thickness  $r_0$ of the axial pinch
is then:
\begin{equation}
r_0^2 = {{E_0^2 - c^4}\over{ c^2 \Omega_0^2}}
\label{ffpinchradius} \end{equation}
Eq.(\ref{solffpinchrelat})
establishes a relation between the total
electric current $I_{\infty}(b)$ flowing through the
polar boundary layer and the limit of $(I/r^2)$ as follows:
\begin{equation}
\lim_{r \rightarrow 0} \left({I \over r^2} \right) =
I_{\infty}(b) \ {{c^2 \Omega_0^2}\over{
E_0^2 - c^4 }}
\label{Ioverr2ff} \end{equation}
For small $I_{\infty}(b)$ the density and magnetic field 
on  the axis are given by:
\begin{equation}
\rho_0(b) = {{\mu_0 \alpha_0^2 c^2}\over{ E_0 \Omega_0}}
\ \lim_{r \rightarrow 0} \left({I \over r^2} \right)
\label{rhoataxisff} \end{equation}
\begin{equation}
B_{P0}^{2} (b)=
{{\mu_0^2 c^2 }\over{\Omega_0^2}} \ {{ E_0^2 - c^4 }\over{E_0^2}} \
\left( \lim_{r \rightarrow 0} \left({I \over r^2} \right) \right)^2
\label{Bpolataxisff} \end{equation}
When the first integrals are almost constant in
the boundary layer, $K_{\infty}(b)$
is, from Eq.\ref{IversusK}:
\begin{equation}
K_{\infty}(b) = {{ c^2 I_{\infty}(b) }\over{E_0}}
\label{IversusKapp} \end{equation}
Using Eqs.(\ref{rhoataxisff}), (\ref{Bpolataxisff}) and
(\ref{IversusKapp}), Eq.(\ref{Ioverr2ff}) reduces to:
\begin{equation}
{{B_{P0}^2(b) }\over{ \mu_0 \rho_0(b)}} 
= {{\Omega_0 K_{\infty}(b) }\over{\alpha_0}}
\label{ffBennetrelat}  \end{equation}
For jets having a force-free polar boundary
layer, this equation replaces Eq.(\ref{BennetrelatK}).
Compared to Eq.(\ref{BennetrelatK}), 
which applies to gas-pressure-supported
polar boundary layers, the axial Alfv\'en velocity has replaced 
in Eq.(\ref{ffBennetrelat}) the axial sound speed.

\section{Null-Surface Boundary Layers}
\label{secnullsurf}

\subsection{Divergence of Mass-to-Magnetic Flux Ratio at Null Surfaces}
\label{subsecdivalpha}

At a null surface of flux parameter $a_n$,
the mass flux to magnetic flux ratio,
$\alpha (a)$, defined by Eq.(\ref{defalpha}),
diverges if there is a mass flux on this null surface.
It has been shown in \citet{HNI} that near $a_n$ the
function $\alpha(a)$ behaves as
\begin{equation}
\alpha(a) \approx {{1}\over{ \vert a_n - a \vert^{\nu}  }}
\label{divergealpha}
 \end{equation}
where $\nu$ is positive and stricly less than unity.
In the case, assumed below, of a field of a dipolar type of symmetry,
the null surfaces reduce to only the equatorial plane.
We now outline the solution
near the equator.

\subsection{Solution in the Equatorial Boundary Layer}
\label{subsecsolequator} 

The structure of the flow is given by 
Eq.(\ref{TFinpolarBL}). We assume that the first integrals
$Q$ and $\Omega$ are almost constant in the boundary
layer, with values $Q_e$ and $\Omega_e$ but we cannot
assume this for $\alpha$, because of its divergence.
Taking into account the vanishing of
$(\vec{\nabla}a \cdot \vec{\nabla}r)$ at
the equator, Eq.(\ref{TFinpolarBL}) becomes:
\begin{equation}
\vec{\nabla}a \cdot \vec{\nabla}
\left(  Q_e \rho^{\Gamma} +
{{ \Omega_e^2 \rho^2 R^2 }\over{ 2 \mu_0 \alpha^2}}
\right)= 0
\label{relassTFequat2} \end{equation}
$R$ enters Eq.(\ref{relassTFequat2}) as a  parameter since 
$(\vec{\nabla}a \cdot \vec{\nabla} R)$ vanishes.
Eq.(\ref{relassTFequat2}) integrates to:
\begin{equation}
Q_e \rho^{\Gamma} + {{\rho^2 \Omega_e^2 R^2}\over{ 2 \mu_0 \alpha^2}} 
= Q_e \rho_e^{\Gamma}(R)
\label{pressequileq} \end{equation}
$\rho_e(R)$ being  the equatorial density at the distance $R$.
The distribution of magnetic flux
in  the equatorial sheet can be found in terms of the parameter
\begin{equation}
X = {{\rho}\over{\rho_e(R) }}
\label{paramequat} \end{equation}
From Eq.(\ref{pressequileq}), 
$\alpha$ can be  expressed at a given $R$ 
as a function of $X$ by:
\begin{equation}
{{1 }\over{\mu_0 \alpha^2(a) }} = {{2 Q_e \rho_e^{\Gamma - 2}(R)}\over{\Omega_e^2 R^2}}
\left({{1 }\over{X^2}} - {{1}\over{ X^{2 - \Gamma} }} \right)
\label{alphaeqparam} \end{equation}
This implicitly determines $a(X)$ since
$\alpha$ is supposedly known as a  function $a$.
The distribution of flux with
latitude angle $\psi$, at almost constant $R$, is infered using 
Eq.(\ref{defalpha}). 
Eqs.(\ref{gradaversuspsi}), (\ref{alphaeqparam}) and
(\ref{defE}) yield  the following differential equation
for $\psi$ at given $R$:
\begin{equation}
R \ d\psi = - {{\alpha(a) da }\over{ R X \rho_e c}}
\ 
{{c^2 + {{2 - \Gamma}\over{\Gamma -1}} Q_e \rho_e^{\Gamma -1} X^{\Gamma - 1} 
+ 2 Q_e \rho_e^{\Gamma -1}  X^{- 1} }\over{ \sqrt{
E^2 -\left( c^2 + {{2 - \Gamma}\over{\Gamma -1}} Q_e \rho_e^{\Gamma -1} X^{\Gamma - 1}
+ 2 Q_e \rho_e^{\Gamma -1}  X^{- 1} \right)^2 }
}}
\label{psiequatbl} \end{equation}
Although not given here, $\psi(X)$ can be determined by 
quadratures 
by using Eq.(\ref{alphaeqparam}) which,
for known, and locally invertible, $\alpha(a)$ provides 
$a$ as a function of $X$.
  
\subsection{Matching the Equatorial Boundary Layer Solution to the Field}
\label{subsecmatchequator}

In the equatorial boundary layer, the solution
is expressed by
Eqs.(\ref{paramequat}), (\ref{alphaeqparam}) 
and (\ref{psiequatbl}). This boundary layer
solution must asymptotically match the field-region solution, expressed in
differential form by equation (\ref{eqdiffpsiofarel}). 
The outer regions of the boundary layer correspond
to small $X$. In this limit,
$X$ can be eliminated by using  Eq.(\ref{alphaeqparam}), so that 
Eq.(\ref{psiequatbl}) reduces to:
\begin{equation}
d \psi = -  {{\alpha}\over{\vert \alpha \vert}} \ {{\Omega_e \ da}\over{ 
\sqrt{2 \mu_0 Q_e R^2  \rho_e^{\Gamma} }  }}
\
{{c^2 + {{\Omega_e}\over{\mu_0 \vert \alpha \vert}} \sqrt{ 2 \mu_0 Q_e R^2 \rho_e^{\Gamma} } 
}\over{
c \ \sqrt{ E^2 
- \left( c^2 + {{\Omega_e}\over{\mu_0 \vert \alpha \vert }} 
\sqrt{ 2 \mu_0 Q_e R^2 \rho_e^{\Gamma} } \ \right)^2  } }} 
\label{psiinouterequaBL}  \end{equation}
Matching requires that equations
(\ref {eqdiffpsiofarel}) and (\ref{psiinouterequaBL}) become identical, 
implying 
\begin{equation}
\mu_0 K(R) = {{\alpha}\over{\vert \alpha \vert}} \sqrt{ 2 \mu_0 Q_e R^2 \rho_e^{\Gamma}(R) }
\label{bennetequator} \end{equation}
This  expresses the balance between the gas pressure force
and the electromagnetic force across the 
equatorial boundary layer. When the proper current, $K(R)$,
approaches a finite limit as $R$ grows to infinity,
the equatorial density 
decreases as 
\begin{equation}
\rho_e(R) \sim R^{- {2\over \Gamma}}
\label{assrhoeqcylind} \end{equation}
For kinetic winds, $K(R)$ decreases as
$(\ln (R/\ell))^{-1}$. The 
equatorial density then declines with distance as
\begin{equation}
\rho_e(R) \sim \left( R \  \ln(R/\ell) \right)^{- {2\over \Gamma}}
\label{assrhoeqparab}  \end{equation}
 
\section{Shape of the Magnetic  Surfaces}
\label{secshape}
 
We have now obtained a complete solution in the asymptotic domain,
both in field regions, near the pole and near null magnetic surfaces. 
The dependence of $K(b)$ on $b$ is obtained from 
Eqs.(\ref{BennetrelatK}) and (\ref{equaforn0}).
We now possess  all the information needed
to calculate the shape of magnetic surfaces 
in any region of the asymptotic domain.

\subsection{Magnetic Surfaces in Field Regions of Poynting Jets}
\label{subsecshapefieldPoynt}

The magnetic surfaces of Poynting jets  
near the polar axis are cylinders.
Their radius is given by Eq.(\ref{rofafarfield}) (with 
Eqs.(\ref{pinchradius}) and (\ref{BennetrelatK})) if they
are in the free field, and 
by equations (\ref{rofxWKB})--(\ref{aofxrelWKB})
if they are in the polar boundary layer.  
The outer-most magnetic surfaces in such flows
are conical. The angle $\psi$ 
defined by Eq.(\ref{coneeq}) becomes independent of
$b$ and is given by Eq.(\ref{psiofarel}).
How cylindrical and conical regions smoothly merge
 one into the other is to be discussed in a 
forthcoming paper \citep{HNIII}.
The shape of magnetic surfaces in the equatorial 
boundary layer of Poynting jets  is discussed 
in section (\ref{subsecshapeequator}).

\subsection{Magnetic Surfaces in Field Regions of Kinetic Winds}
\label{subsecshapefieldWKB}

The shape of poloidal field lines in kinetic winds is described by the
differential system
\begin{eqnarray}
dr = \cos\psi(a, b) \ \ db
\nonumber\\
dz = \sin\psi(a, b) \ \ db
\label{fieldlineeq} \end{eqnarray}
with $\psi(a, b)$ given by Eq.(\ref{psiofarel}), 
supplemented by Eqs. (\ref{KofRexplicit}) or (\ref{chiofaandR}).
In the case where $\psi$ is close to $\pi/2$ we obtain:
\begin{equation}
{{r}\over{\ell}} = {{1}\over{1 - k(a)}}
\left({{2z}\over{\ell}}\right)^{1 - k(a)}
\label{paraboloidshape} \end{equation}
where $k(a)$ is defined by Eq.(\ref{kofa}). 
The magnetic surfaces are 
a collection of nested power-law paraboloids of variable exponent
$q(a) = (1 - k(a))$ as represented in Figure 1.
The shape of magnetic surfaces for which
the approximation $\psi \approx \pi/2$ is invalid  can
be treated as in \citet{HNI}, with a similar result.

\subsection{Magnetic Surfaces in the Polar Boundary Layer}

The solution deep inside the polar boundary layer
of kinetic winds is obtained 
from Eqs.(\ref{rofxWKB})--(\ref{aofxrelWKB}) 
in the limit of $x$ close to unity.
The axial density, $\rho_0(b)$ is given by Eq.(\ref{equaforn0})
with the factor $\lambda$ now being equal to
\begin{equation}
\lambda_0 = \int_0^A \ {{\Omega(a') c^2}\over{\Omega_0 
\sqrt{E^2(a') - c^4} }} \ {{da'}\over{a_0}} 
\label{lambda0}
\end{equation}
Eq.(\ref{rofxWKB}) can be written as
\begin{equation}
{{r^2}\over{\ell^2}} = {{1}\over{n_0^{2 - \Gamma} (b) }}
\ \left( {1 \over x} - {{1}\over{x^{2 - \Gamma} }} \right)
\label{rofxequateur}
\end{equation}
while Eq.(\ref{aofxrelWKB}) provides the value of the parameter $x$
in terms of $a$ by:
\begin{equation}
\left(\ln({1 \over x})
-  {{2 - \Gamma}\over{\Gamma - 1}} (1 - x^{\Gamma -1}) \right)
= {{a}\over{a_0}} \ {{c^2}\over{\sqrt{E_0^2 - c^4}}} \ 
{{1}\over{n_0^{\Gamma - 1} (b) }}
\label{aofxequateur}
\end{equation}
From Eq.(\ref{nzeroofRsimple}), the dimensionless axial density is
in this case given by:
\begin{equation}
{{1}\over{n_0^{\Gamma - 1} (b) }}
= {{\ln \left( {{4 b^2}/{\ell^2}} \right) }\over{ \lambda_0}} 
\label{n0equateur}
\end{equation}
When $x$ is close to unity, it can be eliminated 
between Eqs.(\ref{rofxequateur}) and (\ref{aofxequateur}),
taking $b$ to be almost equal to $z$. This gives the
shape of magnetic surfaces 
in this region as:
\begin{equation}
{r \over \ell} = \sqrt{ {{a}\over{a_0}} } \ 
{{c}\over{(E_0^2 - c^4)^{ {1 \over 4} } }} \ 
\left( {{2 }\over{\lambda_0}} \right)^{ {{1}\over{ 2(\Gamma - 1) }} }
\ \left( \ln \left({{2 z}\over{\ell}} \right) 
\right)^{ {{1}\over{ 2(\Gamma - 1) }} }
\label{shapeincenterofbl}
\end{equation}

\subsection{Magnetic Surfaces in the Equatorial Boundary Layer}
\label{subsecshapeequator}

The information on the shape of magnetic surfaces in the equatorial
boundary layer is contained in the parametric solution provided by
Eqs.(\ref{alphaeqparam})--(\ref{psiequatbl}), the equatorial
density $\rho_e(R)$ being related to $K(R)$ 
by Eq.(\ref{bennetequator}). The latter 
approaches a non-vanishing constant
for Poynting jets and decreases logarithmically for kinetic winds.  It
can therefore be written as
\begin{equation}
K(R) = {{J_m}\over{ \left(ln (R/\ell) \right)^m }}
\label{behaveIinfty} \end{equation}
where $m=0$ for Poynting jets and $m=1$ for kinetic winds.
The factor $J_m$ is different in the two cases.
Eq.(\ref{bennetequator}) 
then gives the associated equatorial density. 
For small $X$, Eqs.(\ref{alphaeqparam})--(\ref{psiequatbl}) 
give the results
for the field-region. In this  limit, 
$\alpha$ can be explicitly obtained
from Eq.(\ref{alphaeqparam})
in terms of $X$ and Eq.(\ref{psiinouterequaBL})
integrates to:
\begin{equation}
{{z}\over{R}} = \int_a^A \   
{{\alpha}\over {\vert \alpha \vert}} \ 
{{\Omega_e \ da}\over{ \sqrt{2 \mu_0 Q_e R^2  \rho_e^{\Gamma}(R) }  }} \
{{c^2 + {{\Omega_e}\over{\mu_0 \vert \alpha \vert}} 
\sqrt{ 2 \mu_0 Q_e R^2 \rho_e^{\Gamma}(R) }
}\over{
c \ \sqrt{ E^2 - \left( c^2 + {{\Omega_e}\over{\mu_0 \vert \alpha \vert }}
\sqrt{ 2 \mu_0 Q_e R^2 \rho_e^{\Gamma}(R) } \ \right)^2  } }}
\label{zofRequatoroutskirt} \end{equation}
For Poynting jets, $K(R)$ approaches a finite constant
as does $R^2 \rho_e^{\Gamma}(R)$ (from Eq. (\ref{bennetequator})). 
Consequently,
the magnetic surfaces become conical at the outskirts of the
equatorial boundary layer.
For kinetic winds $K(R)$ decreases logarithmically 
and the equatorial density declines as 
$ ( R \  ln(R/\ell))^{- {2\over \Gamma}}$.
Eq.(\ref{zofRequatoroutskirt}) shows that in the equatorial
boundary layer $z$ is proportional to $(A -a) \  R \ln (R/\ell) $.
The magnetic surfaces at the outskirts of
the equatorial boundary layer become slightly convex paraboloids.
By contrast, deep inside the equatorial boundary layer,
gas pressure dominates, $X$ is close to unity and
Eqs.(\ref{alphaeqparam}) and (\ref{psiequatbl}) give:
\begin{equation}
z = \int_a^A \ {{ \alpha(a) \ da}\over{ R \ \rho_e(R) }}
\ {{c^2 + {{\Gamma}\over{\Gamma -1}} Q_e \rho_e^{\Gamma}(R) }\over{
c \ \sqrt{ E^2
- \left( c^2 + {{\Gamma}\over{\Gamma -1}} Q_e \rho_e^{\Gamma}(R) \ \right)^2  } }}
\label{zofRequatorclose} \end{equation}
In this region $z(R)$ scales, at fixed $a$, as $z \sim R^{ {{2 -
\Gamma}\over{\Gamma}} }$ for Poynting jets and as $ z \sim R^{ {{2 -
\Gamma}\over{\Gamma}} } \ \left(ln {{R}\over{\ell}} \right)^{2 \over
\Gamma}$ for kinetic winds.  Magnetic surfaces are concave, i.e. they
bend towards the equator.  This result is quite consistent with the
properties of such flows discussed by 
\citet{Okamoto99, Okamoto00, Okamoto01}
and \citet{BeskinOkamoto00}.  It
does not contradict our earlier results that the poloidal field lines
cannot asymptotically bend towards the equator \citep{HN89}.  Actually,
any such poloidal line eventually escapes 
the equatorial boundary layer region
at a finite distance, rejoining the field region where it becomes
convex, bending away from the equator.  It can be shown that when
poloidal field lines exit the equatorial boundary layer, they do so at
an angle to the equatorial plane that decreases with distance. The
proof is similar to the non-relativistic case \citep{HNI}.

\section{Conclusions}
\label{secconclu}

\noindent
Our main results are summarized as follows:

\noindent
I. We have derived a global solution for the asymptotic structure of
relativistic, stationary, axisymmetric, polytropic, unconfined,
perfect MHD winds, assuming their five lagrangian first integrals to
be known as a function of the flux parameter, $a$. Current-carrying
boundary layers along the polar axis and at null magnetic surfaces are
features of this solution, which is given in the form of matched
asymptotic solutions separately valid inside and outside these boundary
layers.

\noindent
The asymptotic structure consists of field regions where, although the
electromagnetic force decreases to zero asymptotically, it still
dominates the force balance. In these field-regions the magnetic field
structure is shown to be described by the Hamilton-Jacobi equation
(\ref{HamJacobSrel}) which we solved.  We obtained the distribution of
flux as represented by equations (\ref{psiofarel}) and
(\ref{rofafarfield}). The poloidal 
proper current $K$ (Eq.(\ref{KintermsofE})) is shown to
remain constant in each of these field-region cells, 
approaching
a value independent of $a$.  For a non-vanishing asymptotic value of
$K$ a substantial part of the energy reaches the asymptotic domain in
electromagnetic form. For vanishing $K$, all of the wind energy is in
kinetic form.  We found that $K$ 
may approach zero as the inverse
 logarithm of distance.

\noindent
Field regions are bordered by 
\begin{enumerate}
\item{boundary layers along the polar axis and }
\item{null surface boundary layers }
\end{enumerate}
These boundary layers have the structure of charged column or sheet
pinches supported by plasma pressure. For very cold winds, the support
is by magnetic poloidal pressure.  They carry the poloidal electric
current still present in the asymptotic domain.

\noindent
II. The polar boundary layer has been shown to have the structure of a
charged column-pinch for which a Bennet relation between axial
pressure and proper axial current is given by Eq.(\ref{BennetrelatK}).
The distribution of flux with radius in this region is explicitly
given by Eqs.(\ref{rofxWKB})--(\ref{aofxrelWKB}).

\noindent
III. The equatorial region, or more generally the vicinity of any null
magnetic surface, has been shown in section (\ref{secnullsurf}) to have the
structure of a charged sheet-pinch. The density in the equatorial
boundary layer of a Poynting jet decreases as described by
Eq.(\ref{assrhoeqcylind}), while in the equatorial boundary layer of a
kinetic wind it decreases as described by Eq.(\ref{assrhoeqparab}).

\noindent
IV. We have derived solutions valid in these regions and have matched
them asymptotically to the field region solutions.  Matching between
the polar boundary layer and the field-regions determines the decline
with distance of the density at the polar axis or, equivalently, the
proper current around the polar axis.  The variation with distance of
the axial density is given by the solution of Eq.(\ref{equaforn0}),
considering Eqs.(\ref{equaforrho0}),(\ref{scalepinchalf}),
(\ref{normrhotoalfdens}),(\ref{referenceflux}) and (\ref{deflambda}).
A simplified form of Eq.(\ref{equaforn0}) is
Eq.(\ref{nzeroofRsimple}).

\noindent
V.  The geometry of magnetic surfaces in all parts of the asymptotic
domain has been explicitly deduced in terms of the first-integrals in
section (\ref{secshape}) for 
both kinetic winds and Poynting jets regimes.
In particular, the magnetic
surfaces in field-regions of a relativistic kinetic wind are
paraboloids with a variable exponent, given by
Eqs.(\ref{paraboloidshape}) and (\ref{kofa}), while, deep in the polar
boundary layer, magnetic surfaces have exponential shapes, as
represented by Eq.(\ref{shapeincenterofbl}).  Deep in the equatorial
boundary layer, the shape of magnetic surfaces is represented by
equation (\ref{zofRequatorclose}). Note that any magnetic surface
embedded in this equatorial boundary layer eventually finds its way
out at some finite distance.

\begin{figure}
\plotone{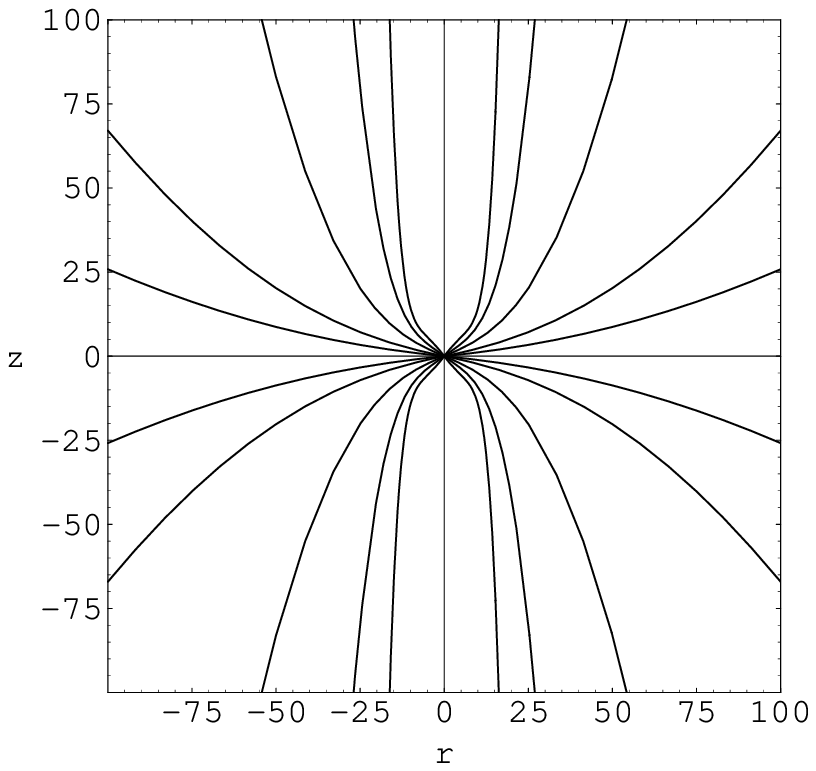}
\caption{The magnetic field structure for kinetic winds 
in the asymptotic field region, from Eq.(\ref{paraboloidshape}).  The
functions $E(a)$ and $\Omega(a)$ have been taken as constant which
implies that $q(a) = (a/A)$. On the pole $q(0)$ = 0 and at the equator
$q(A)$ = 1.  The field lines in each quadrant correspond to $a/A = $
$0.2$, $0.4$, $0.6$, $0.8$, $0.9$. An interpolation
formula has been used to connect the asymptotic solution to a
split-monopole field at the origin. The scale for $r$ and $z$ is
arbitrary.
\label{fig1}}
\end{figure}

\begin{figure}
\plotone{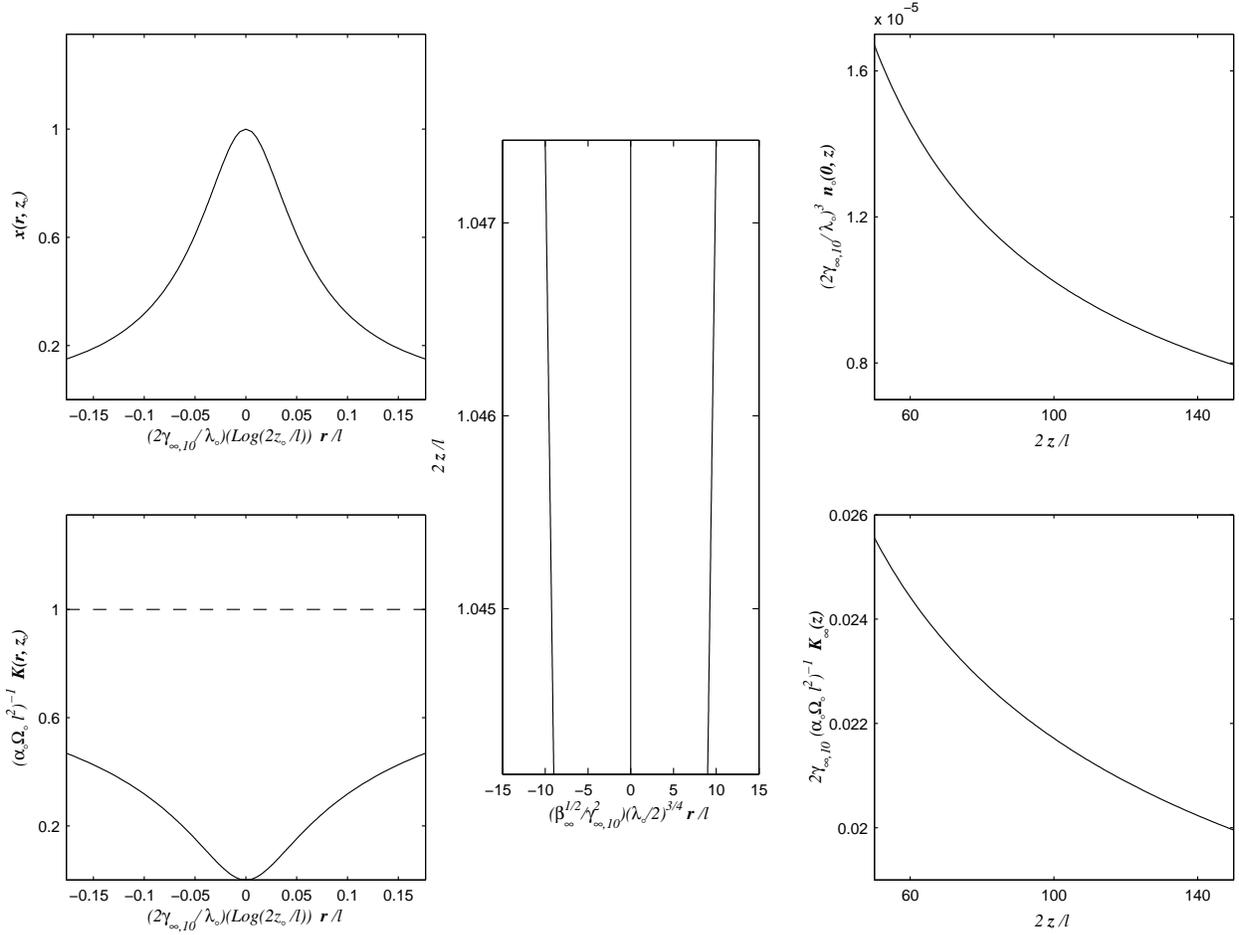}
\caption{The behaviour of kinetic winds near the polar axis. The panels show: magnetic field
structure (central frame); the normalized density (Eq.(\ref{normrhotoalfdens})) at the polar axis vs. the distance
along the polar axis (upper right); the total proper current integrated about the polar axis vs. the 
distance along the polar axis (lower right); the ratio of the density to its polar value at
the same $z$ across the polar pinch (upper left) and the integrated proper current across the polar
axis (lower left). The latter two quantities are plotted for some arbitrarily chosen $z_0$.
In the central frame, $a$ = $a_0\left(\frac{\lambda_0}{2}\right)^{3/2}$ (Eq.(\ref{shapeincenterofbl})).
The slightly parabolic shape of the field lines is not clearly visible on the
scale of the plot. The right panel curves are from Eqs.(\ref{BennetrelatK}), (\ref{scalepinchalf}) -- (\ref{referenceflux}) and (\ref{nzeroofRsimple}) with $\lambda_0$ given
by Eq.(\ref{deflambda}) for negligibly small $K_\infty$. The left panel 
curves are from Eqs.(\ref{definexcylindr})--(\ref{aofxrelWKB}), 
(\ref{rBthetaass}), (\ref{K2asfunctionofI2}),
(\ref{scalepinchalf})--(\ref{referenceflux}) and (\ref{nzeroofRsimple}).
\label{fig2}}
\end{figure}

\begin{figure}
\plotone{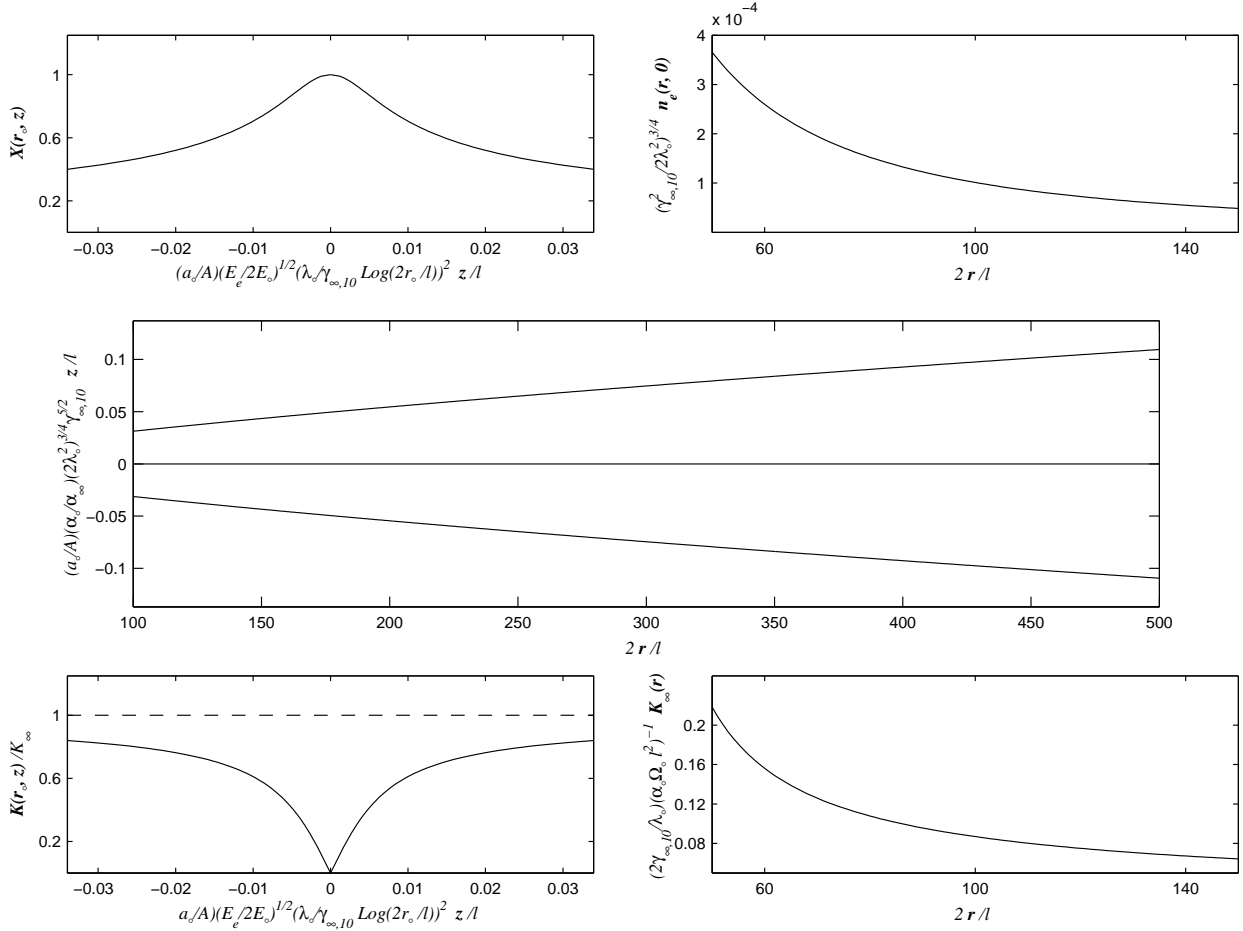}
\caption{The behaviour of kinetic winds about the equatorial boundary layer. The panels show:
the magnetic field structure (central); the normalized density $n_e$ = $\frac{\rho(r,0)}{\rho_{A0}}$
in the equatorial plane vs. distance along the equatorial plane (upper right); the total 
proper current integrated about the equatorial plane vs. distance along the equatorial plane (lower
right); the ratio of the density to its equatorial value at the same $r$ across the equatorial
sheet (upper right) and the integrated proper current across the equatorial sheet (lower left). The
latter two quantities are plotted for some arbitrary value of $r$, $r_0$. For simplicity,
$\Omega_e$, $\alpha_e$ and $Q_e$ have been set equal to $\Omega_0$, $\alpha_0$ and $Q_0$
respectively. The field lines (central frame, Eq.(\ref{zofRequatorclose})) correspond to $\frac{a}{A}$ = 0.9.
Their parabolic shape is not clearly visible on the scale of the plot. The plots in the
right panel are from Eqs.(\ref{bennetequator}) and (\ref{assrhoeqparab}), while those in the left panels are from Eqs.(\ref{divergealpha}), (\ref{paramequat}),
(\ref{alphaeqparam}) and (\ref{rBthetaass}) with $\nu$ = $\frac{1}{2}$.
\label{fig3}}
\end{figure}

\begin{figure}
\epsscale{0.85}
\plotone{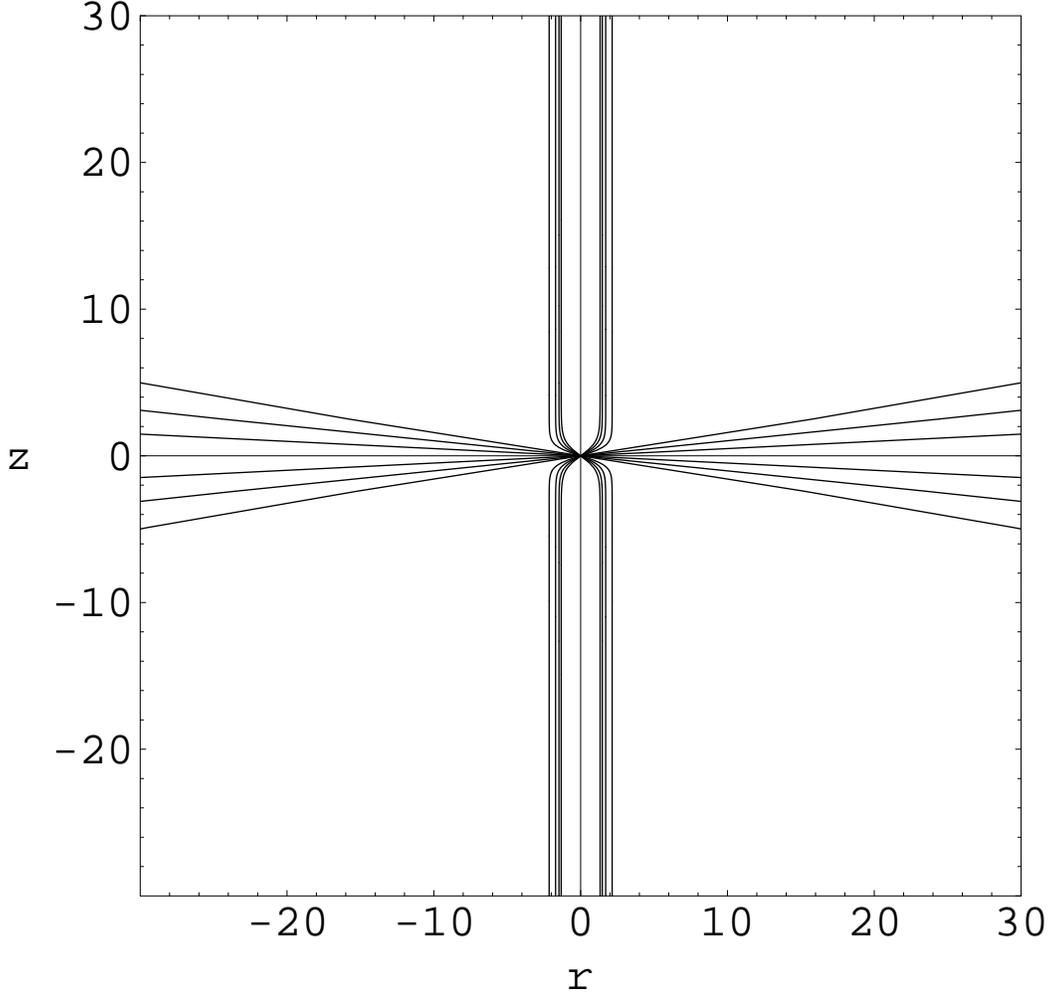}
\caption{The magnetic field structure for Poynting 
jets in the field region, from Eqs.(\ref{psiofarel}) and
(\ref{rofafarfield}). From pole to equator, the contours are for
$(a/A) =$ $0.4$, $0.6$, $0.7$, $0.85$, $0.9$ and $0.95$.  $E(a)$ and
$\Omega(a)$ are constants, while we assume, as discussed in Paper III,
that $\alpha(a) = (K \Omega /(E- c^2)) 
\lbrace 1 + \left[((a/A) -0.8)^2 /\sqrt{1 -(a/A)}\right] \rbrace$.
The unit of the plot for the variables $r$ and $z$ is the scale $\ell$
defined in Eq.(\ref{scalepinchalf}).  An interpolation formula has
been used to connect the asymptotic structure of the field to a split
monopole field near the origin.  Eqs.(\ref{psiofarel}) and
(\ref{rofafarfield}) become increasingly accurate with larger $r/\ell$
and $z/\ell$, the scale of the transition between the split monopole
structure near the origin and the asymptotic one being arbitrary.
\label{fig4}}
\end{figure}

\begin{figure}
\epsscale{0.85}
\plotone{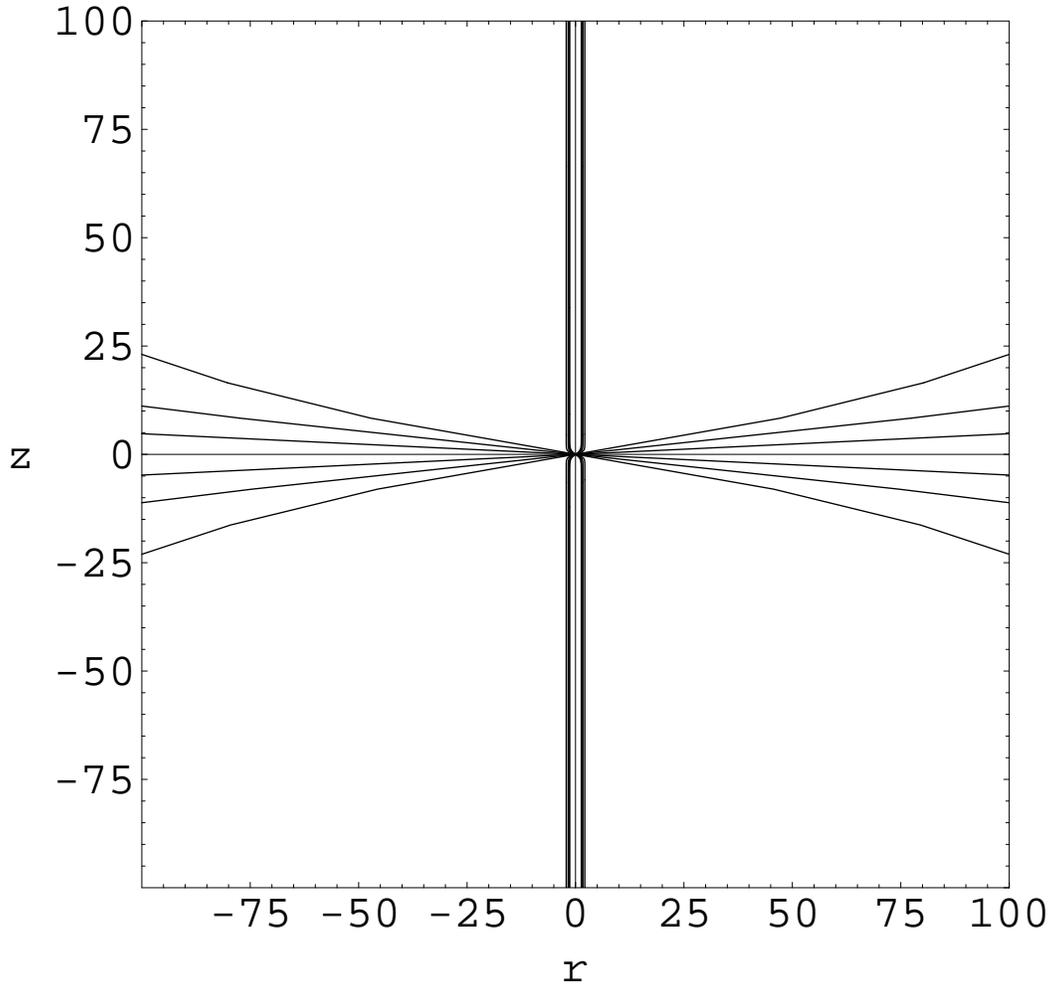}
\caption{ Same as figure (\ref{fig4}). The scale in $(r/\ell)$ and $(z/\ell)$
is larger.
\label{fig5}}
\end{figure}

\begin{figure}
\plotone{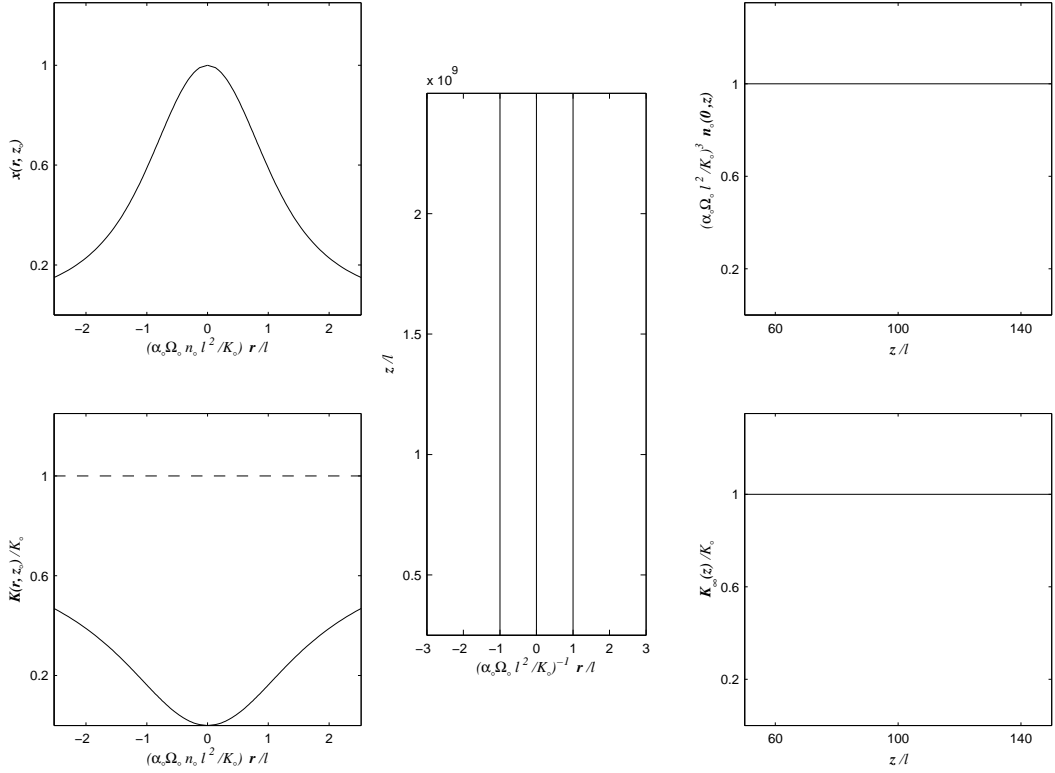}
\caption{The behaviour of Poynting jets near the polar axis. The plots show: magnetic field
structure (central frame); the normalized density at the polar axis vs. the distance along
the polar axis (upper right); the total proper current integrated about the polar axis vs. the 
distance along the polar axis (lower right); the ratio of the density to its polar value 
across the polar pinch (upper left) and the integrated proper current across the polar axis (lower
left). The latter two quantities are plotted for some arbitrarily chosen $z_0$. The central
frame shows that the field lines are exact cylinders 
(Eq.(\ref{pinchradius})). The right panel curves use Eqs.(\ref{BennetrelatK}) 
and (\ref{normrhotoalfdens}). The left panel curves are obtained 
from Eqs.(\ref{definexcylindr})--(\ref{aofxrelWKB}), (\ref{rBthetaass}) and (\ref{K2asfunctionofI2}). $K_0$
here is the absolute minimum of $\frac{\alpha E}{\Omega}$.
\label{fig6}}
\end{figure}

\begin{figure}
\plotone{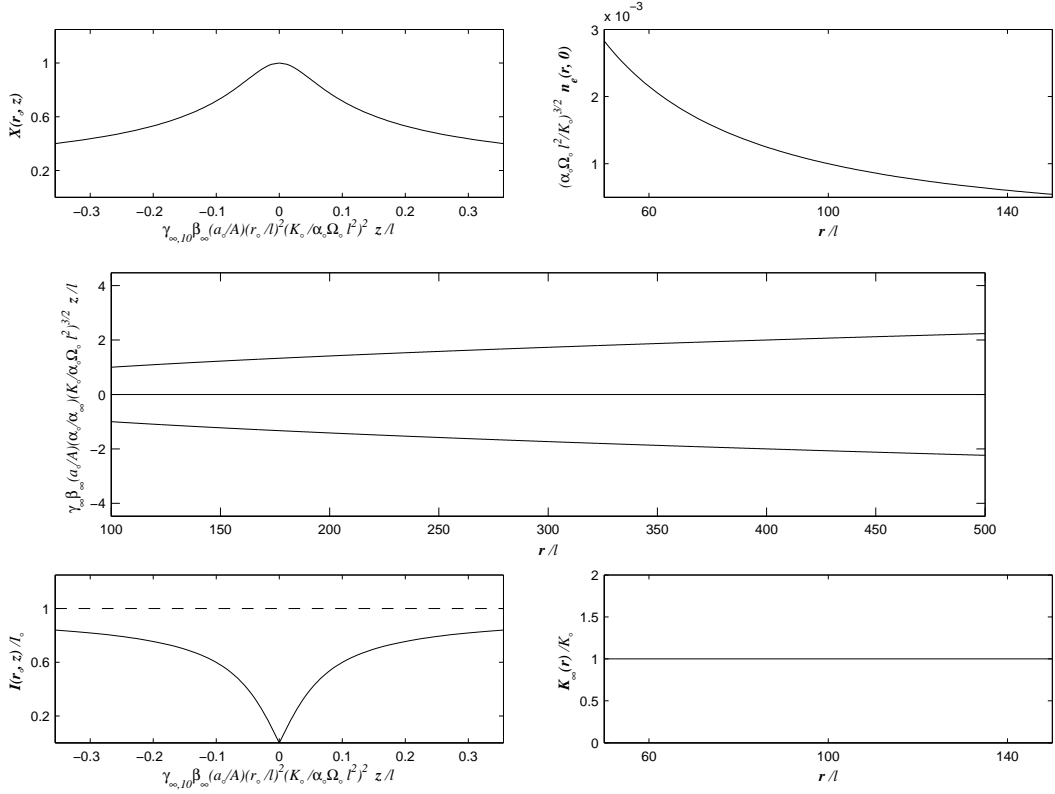}
\caption{The behaviour of Poynting jets about the equatorial boundary layer. The plots show: 
magnetic field structure (central frame, from Eq.(\ref{zofRequatorclose})); the normalized density $n_e$ = 
$\frac{\rho(r,0)}{\rho_{A0}}$ in the equatorial plane vs. the distance along the equatorial
plane (upper right); the total proper current integrated about the equatorial plane vs. the distance 
along the equatorial plane (lower right); the ratio of the density to its equatorial value
at the same $r$ across the equatorial sheet (upper left) and the integrated proper current across
the equatorial sheet (lower left). The latter two quantities are plotted for some 
arbitrary value of $r$, $r_0$. The field structure (central frame) is plotted for
$\frac{a}{A}$ = 0.9. The plots in the right panels are from Eqs.(\ref{bennetequator}) and (\ref{assrhoeqcylind}), while those in
the left panels are from Eqs.(\ref{paramequat}), (\ref{alphaeqparam}), (\ref{rBthetaass}) and (\ref{divergealpha}) with $\nu$ = $\frac{1}{2}$.
\label{fig7}}
\end{figure}

These results are illustrated in Figures \ref{fig1}-\ref{fig7}.
Figures \ref{fig1}-\ref{fig3} represent properties of 
kinetic winds as deduced from our analysis. Figure \ref{fig1}
illustrates the geometry of magnetic surfaces. Figures \ref{fig2}
and \ref{fig3} represent the structure of the polar and equatorial
boundary layers. 
Figures \ref{fig4}-\ref{fig7} represent similar 
properties for Poynting jets.

In all cases, the polar and null surface boundary layers which carry
residual electric current may stand out observationally against the
field-regions, both because of their large density contrast to them
and because they possess a source of free energy that makes them
potentially active by the development of instabilities. It is
noteworthy that the density about the pole either does not decline
with distance in the case of Poynting jets, or only very slowly (as a
negative power of the logarithm of the distance) in the case of
kinetic winds.  It may be that what is observed as a jet be but the
dense and active polar boundary layer of a flow developed on a much
larger angular scale. This diffuse flow may be itself barely visible
because of its very low density and activity. Null surface boundary
layers, for example equatorial ones, enjoy a more favourable status
than field regions from this point of view \citep{BeskinOkamoto00},
and may be observed in association with jets.  The X-ray structure of
the Crab Nebula \citep{Weisskopfetal00} can be understood in such a
framework \citep{blandford2002}. In addition, our work is relevant to
the large scale aspects of pulsar winds, jets from active galaxies and
gamma ray bursts \citep{vlahakis1,vlahakis2}. It is interesting to
note that, from our analysis, the highest degrees of collimation are
associated with flows which carry significant Poynting flux. In our
next paper \citep{HNIII}, we show that such flows should actually
occur, owing to the slow decline of the asymptotic proper current with
distance.

\acknowledgements

The authors thank the Space Telescope Science Institute and the Johns
Hopkins University for continued support to their collaboration.  JH
also thanks the EC Platon program (HPRN-CT-2000-00153) and the Platon
collaboration. CN is pleased to thank the Director of ESO for support
and hospitality during which time this paper was completed.  We thank
Sundar Srinivasan for significant help with the figures.

\newpage


\clearpage

\clearpage

\clearpage

\clearpage

\clearpage

\clearpage

\clearpage


\bigskip
\bigskip


\begin{thebibliography}{}
\bibitem[Appl and Camenzind(1993a)]{Appl93a}
Appl, S. and Camenzind, M. 1993 \aap 270, 71
\bibitem[Appl and Camenzind(1993b)]{Appl93b}
Appl, S. and Camenzind, M. 1993 \aap 274, 699.
\bibitem[Ardavan(1979)]{Ardavan} Ardavan, H.  1979, \mnras 189, 341.
\bibitem[Begelman \& Li(1994)]{BegelmanLi94} Begelman, M. C. and Li, Z. Y. 
1994 \apj 426, 269.
\bibitem[Bell \& Lu\c cek(1995)]{BellLucek}
Bell, A.R. and Lu\c cek,  S.G. 1995, \mnras 277, 1327.
\bibitem[Beskin et al.(1993)]{Beskinbook} 
Beskin, V.S., Gurevich, A.V. and Istomin, Ya. N., 1993
Physics of the pulsar magnetosphere,
Cambridge University Press.
\bibitem[Beskin et al.(1998)]{Beskinetal98}
Beskin,  V.S., Kuznetsova, I.V., and Rafikov, R.R. 1998 \mnras 299, 341
\bibitem[Beskin \& Malyshkin(2000)]{BeskinMalyshkin00}
Beskin,  V.S. and Malyshkin, L.M. 2000 Astronomy Letters 26, 208.
\bibitem[Beskin \& Okamoto(2000)]{BeskinOkamoto00}
Beskin,  V.S. and Okamoto, I. 2000 \mnras 313, 445.
\bibitem[Beskin \& Par'ev(1993)]{BeskinParev}
Beskin,  V.S. and Par'ev V. 1993 Uspekhi 36, 529
\bibitem[Blandford(2002)]{blandford2002} Blandford, R.D. 2002, in Lighthouses of
the Universe, ed. M. Gilfanov et al. (Springer:Berlin), p. 381
\bibitem[Blandford \& Payne(1982)]{BlandfordPayne}
Blandford, R. D. and Payne, D. G., 1982  \mnras 199, 88
\bibitem[Bogovalov(1999)]{bogovalov99} Bogovalov, S.V. 1999 \aap 349, 1071
\bibitem[Bogovalov(2001)]{bogovalov01} Bogovalov, S.V. 2001 \aap 371, 1155
\bibitem[Bogovalov \& Tsinganos(1999)]{BogovalovTsingan99}
Bogovalov, S. and Tsinganos, K. 1999 \mnras 305, 211.
\bibitem[Bogovalov \& Tsinganos(2001)]{TsinganBogovalov01}
Bogovalov, S. and Tsinganos, K.  2001 \mnras 325, 249
\bibitem[Camenzind(1986a)]{Camenzind86a}
Camenzind, M., 1986 \aap 156, 137.
\bibitem[Camenzind(1986b)]{Camenzind86b}
Camenzind, M., 1986 \aap 162, 32.
\bibitem[Camenzind(1987)]{Camenzind87}
Camenzind, M. 1987 \aap  184, 341.
\bibitem[Camenzind(1989)]{Camenzind89}
Camenzind, M. 1989 in Accretion Disks and Magnetic Fields in Astrophysics,
ed. G. Belvedere, (Dordrecht Kluwer) p 129.
\bibitem[Chiueh et al.(1991)]{ChiuehLiBegelman}
Chiueh, T. Li, Z.Y. and Begelman, M. 1991 \apj 377, 462.
\bibitem[Chiueh et al.(1998)]{Chiuehetal98}
Chiueh, T. Li, Z.Y. and Begelman, M. 1998 \apj 505, 835.
\bibitem[Contopoulos(1994)]{Contopoulos94}
Contopoulos, J. 1994 \apj 432, 508.
\bibitem[Contopoulos(1995)]{Contopoulos95}
Contopoulos, J. 1995 \apj 446, 67.
\bibitem[Contopoulos et al.(1999)]{ContopoulosKazanasFendt}
Contopoulos, I, Kazanas, D. and Fendt, C. 1999, \apj 511, 351.
\bibitem[Contopoulos \& Lovelace(1994)]{ContopoulosLovelace94}
Contopoulos, I. and Lovelace, R.V.E. 1994, \apj 429, 139.
\bibitem[Fendt et al.(1993)]{FendtCamAppl}
Fendt,C.,  Camenzind, M. and  Appl, S. 1993 \aap 300, 791
\bibitem[Fendt \& Camenzind(1996)]{FendtCam96}
Fendt,C. and  Camenzind, M. 1996 \aap 313, 591
\bibitem[Goldreich \&  Julian(1970)]{GJ}
Goldreich, P. and Julian, W. H., 1970 \apj 160, 971.
\bibitem[Heinemann \& Olbert(1978)]{HeinemannOlbert}
Heinemann, M. and Olbert, S., 1978  \jgr  82, 23.
\bibitem[Heyvaerts \& Norman(1989)]{HN89}
Heyvaerts, J. and Norman, C.A. 1989 \apj 347, 1055
\bibitem[Heyvaerts \& Norman(2002a)]{HNI}
Heyvaerts, J. and Norman, C.A. 2002a \apj, submitted
\bibitem[Heyvaerts \& Norman(2002b)]{HNIII}
Heyvaerts, J. and Norman, C.A. 2002b \apj, submitted
\bibitem[Koide et al.(1999)]{Koideetal99}
Koide, S. Shibata, K. and Kudoh, T. 1999 \apj 522, 727
\bibitem[Koide et al.(2000)]{Koideetal00}
Koide, S., Meier, D.L., Shibata, K. and Kudoh, T. 2000 \apj 536, 668
\bibitem[Koide et al.(2002)]{koide} Koide, S., Shibata, K., Kudoh, T. \& Meier, D.L. 2002, Science, {\bf 295}, 1688 
\bibitem[Krasnopolsky et al.(1999)]{Krasnopolskyetal}
Krasnopolsky, R., Li, Z.Y., and Blandford, R. 1999 \apj 526, 631.
\bibitem[Kudoh et al.(1998)]{Kudohetal98}
Kudoh, T., Matsumoto, R., Shibata, K. 1998 \apj 508, 186.
\bibitem[Li(1993a)]{LiThesis}
Li, Z.Y. 1993, PhD Thesis, University of Colorado, Boulder.
\bibitem[Li(1993b)]{Li93b} Li, Z. Y. 1993 \apj 415, 118
\bibitem[Li et al.(1992)]{Lietal92}
Li Z. Y., Chiueh, T. and Begelman, M. 1992 \apj 394, 459.
\bibitem[Lovelace(1976)]{Lovelace76}
Lovelace, R.V.E. 1976  Nature  262, 649
\bibitem[Lovelace et al.(1987)]{Lovelaceetal87}
Lovelace, R.V.E, Wang, J.C.L \& Sultanen, M.E. 1987 \apjl 315, L504 
\bibitem[Lovelace et al.(1991)]{Lovelaceetal91}
Lovelace, R.V.E, Berk, H.L. and Contopolos, J. 1991 \apj 379, 696.
\bibitem[Lovelace et al.(1993)]{Lovelaceetal93}
Lovelace, R.V.E. Romanova, M. and Contopolos, J. 1993 \apj 403, 158.
\bibitem[Lu\c cek \& Bell(1997)]{LucekBell97}
Lu\c cek, S.G.  and Bell, A.R.  1997 \mnras 290, 327
\bibitem[Matsumoto et al.(1996)]{Matsumotoetal96}
Matsumoto, R., Uchida, Y., Hirose, S., Shibata, K.,
Hayashi, M.R., Ferrari, A., Bodo, G \& Norman, C.A. 1996 \apj 461, 115
\bibitem[Michel(1969)]{Michel69}
Michel, F.C. 1969 \apj 158, 727. 
\bibitem[Mobarry \& Lovelace(1986)]{MobarryLovelace}
Mobarry, C.M. and Lovelace, R.V.E. 1986, \apj 309, 455.
\bibitem[Nishikawa et al.(1997)]{Nishikawaetal97}
Nishikawa,  K-I, Koide, S.,  Sakai, J-I,
Christodoulou, D.M., Sol, H. and Mutel, R. 1997 \apj 483, L45
\bibitem[Nitta(1995)]{Nitta95}
Nitta, S-Y 1995 \mnras 276, 825.
\bibitem[Nitta(1997)]{Nitta97}
Nitta, S-Y 1997 \mnras 284, 899.
\bibitem[Nitta et al.(1991)]{Nittaetal91}
Nitta, S-Y., Takahashi, M. and Tomimatsu, A. 1991
Phys. Rev. D, 44, 2295.
\bibitem[Okamoto(1975)]{Okamoto75}
Okamoto, I., 1975  \mnras 173, 357.
\bibitem[Okamoto(1999)]{Okamoto99}
Okamoto, I., 1999 \mnras 307 253
\bibitem[Okamoto(2000)]{Okamoto00}
Okamoto, I., 2000 \mnras 318 250
\bibitem[Okamoto(2001)]{Okamoto01}
Okamoto, I., 2001 \mnras 327, 55
\bibitem[Spruit et al.(1997)]{Spruitetal97}
Spruit, H.C., Foglizzo, T. and Stehle, R. 1997 \mnras 288, 333.
\bibitem[Takahashi et al.(1990)]{Takahashietal90}
Takahashi, M., Nitta, S., Tatematsu, Y. and Tomimatsu, A. 1990
\apj 363, 206.
\bibitem[Tomimatsu(1994)]{Tomimatsu94}
Tomimatsu, A. 1994 \pasj 46, 123.
\bibitem[Tomimatsu \& Takahashi(2003)]{tomimatsu} Tomimatsu, A. \& Takahashi, M. 2003 \apj 592 321
\bibitem[Tsinganos \& Bogovalov(1999)]{TsinganBogovalov99}Tsinganos, K. and Bogovalov, S. 1999 \mnras 305, 211.
\bibitem[Ustyugova et al.(2000)]{Ustyugovaetal00}
Ustyugova G. V., Lovelace, R. V. E., Romanova, M. M., Li, H. and Colgate, S.A. 2000
\apj 541 L21 
\bibitem[Van Putten(1997)]{VanPutten97}
Van Putten,  M.H.P.M. 1997 \apj 488, 69L.
\bibitem[Vlahakis \& K\"onigl(2003a)]{vlahakis1} Vlahakis, N. \& K\"onigl, A. 2003, astro-ph/0303482
\bibitem[Vlahakis \& K\"onigl(2003b)]{vlahakis2} Vlahakis, N. \& K\"onigl, A. 2003, astro-ph/0303483
\bibitem[Weisskopf et al.(2000)]{Weisskopfetal00}
Weisskopf, M.C., Hester, J.J., Tennant A.A.,
Elsner, R.F., Schulz, N.S., Marshall, H.L., Karovska, M.,
Nichols, J.S., Swartz, D.A., Kolodziejczak, J.J. and O'Dell, S.L. 
2000 \apjl 536, L81 

\end{thebibliography}
\end{document}